\documentclass[sigconf]{acmart}

\AtBeginDocument{%
  \providecommand\BibTeX{{%
    \normalfont B\kern-0.5em{\scshape i\kern-0.25em b}\kern-0.8em\TeX}}}

\copyrightyear{2022} 
\acmYear{2022} 
\setcopyright{rightsretained} 

\acmConference[AIES'22]{Proceedings of the 2022 AAAI/ACM Conference on AI, Ethics, and Society}{August 1--3, 2022}{Oxford, United Kingdom}
\acmBooktitle{Proceedings of the 2022 AAAI/ACM Conference on AI, Ethics, and Society (AIES'22), August 1--3, 2022, Oxford, United Kingdom}\acmDOI{10.1145/3514094.3534138}
\acmISBN{978-1-4503-9247-1/22/08}

\usepackage{etoolbox}
\makeatletter
\patchcmd{\maketitle}{\@copyrightpermission}{
   \begin{minipage}{0.3\columnwidth}
     \href{url=https://creativecommons.org/licenses/by/4.0/}{\includegraphics[width=0.90\textwidth]{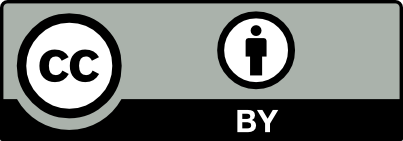}}
   \end{minipage}\hfill
   \begin{minipage}{0.7\columnwidth}
     \href{url=https://creativecommons.org/licenses/by/4.0/}{This work is licensed under a Creative Commons Attribution International 4.0 License.}
   \end{minipage}
  
   \vspace{5pt}
}{}{}

\begin{document}

\title{Artificial Concepts of Artificial Intelligence: Institutional Compliance and Resistance in AI Startups}


\author{Amy A. Winecoff}
\email{aw0934@princeton.edu}
\affiliation{%
  \institution{Princeton University}
  \streetaddress{Sherrerd Hall}
  \city{Princeton}
  \state{New Jersey}
  \country{USA}
  \postcode{08544}
}

\author{Elizabeth Anne Watkins}
\email{ew4582@princeton.edu}
\affiliation{%
  \institution{Princeton University}
  \streetaddress{Sherrerd Hall}
  \city{Princeton}
  \state{New Jersey}
  \country{USA}
  \postcode{08544}
}




\begin{abstract}
  Scholars and industry practitioners have debated how to best develop interventions for ethical artificial intelligence (AI). Such interventions recommend that companies building and using AI tools change their technical practices, but fail to wrangle with critical questions about the organizational and institutional context in which AI is developed. In this paper, we contribute descriptive research around the life of ”AI” as a discursive concept and organizational practice in an understudied sphere--emerging AI startups--and with a focus on extra-organizational pressures faced by entrepreneurs. Leveraging a theoretical lens for how organizations change, we conducted semi-structured interviews with 23 entrepreneurs working at early-stage AI startups. We find that actors within startups both conform to and resist institutional pressures. Our analysis identifies a central tension for AI entrepreneurs: they often valued scientific integrity and methodological rigor; however, influential external stakeholders either lacked the technical knowledge to appreciate entrepreneurs' emphasis on rigor or were more focused on business priorities. As a result, entrepreneurs adopted hyped marketing messages about AI that diverged from their scientific values, but attempted to preserve their legitimacy internally. Institutional pressures and organizational constraints also influenced entrepreneurs' modeling practices and their response to actual or impending regulation. We conclude with a discussion for how such pressures could be used as leverage for effective interventions towards building ethical AI. 
\end{abstract}






\maketitle

\section{Introduction}
Academic researchers, advocacy groups, and technology companies have created guidelines and tools for developing ethical artificial intelligence (AI) \cite{morley2020initial,fish2021reflexive}. This research is intended to ameliorate the considerable negative social impacts produced in AI systems, such as how AI models encode racial and gender biases \cite{caliskan2017semantics, noble2018algorithms, sweeney2013, rekabsaz2020neural, bolukbasi2016man}, worsen disordered eating and body dysmorphia \cite{karizat2021algorithmic}, and magnify inequality \cite{obermeyer2019dissecting, dastin2018, angwin2016}. However, the real world-utility of available interventions for ethical AI remains unclear. 

As with any research intended for real-world applications, robust consideration of the context of implementation is critical. An emerging body of research has begun to recognize that effective change demands non-technical strategies to contend with organizational context \cite{rakova2021responsible}, such as the conditions inside technology firms which might influence, or even prevent, the effectiveness of interventions for ethical AI. Tight development timelines, lack of formal organizational processes, and challenging internal stakeholder dynamics shape how real-world companies can move in the direction of more ethical AI development \cite{holstein2019improving, madaio2020co, madaio2021assessing, rakova2021responsible, hopkins2021machine}. Studies of AI ethics in organizational contexts often focus on interventions such as model fairness  \cite{holstein2019improving, madaio2020co, madaio2021assessing} and model interpretability techniques \cite{bhatt2020explainable, hong2020human, kaur2020interpreting}. However, such studies have largely been constrained to organizations that are mature enough to consider specific AI ethical interventions to begin with. 

AI startups constitute a growing portion of the technology sector \cite{tricot2021}. As a result, these companies and the ethical practices they embrace are likely to play a significant role in the impact of future technology on society. Only a handful of studies have characterized ethical AI development at smaller firms \cite{vakkuri2020just, hopkins2021machine}. This emerging area of research has begun to illuminate the unique challenges nascent firms must address when attempting to adopt responsible, transparent, and accountable AI practices. For example, as with more mature companies, small firms must navigate complex dynamics amongst stakeholders like clients, investors, and regulators but unlike more established organizations, they must do so under significant resource constraints that threaten their very existence \cite{hopkins2021machine}. Therefore, the ethical AI practices they are able to adopt are necessarily limited.  

While existing research has illustrated the organizational constraints to ethical AI, especially intra-organizational dynamics, less is known about how the inter-organizational or field-level dynamics shape firms' capacity to develop ethical approaches. The field-level, i.e., "institutional" dynamics and market-based pressures that impact an organization's chances of survival inevitably alter the structures and practices firms adopt \cite{meyer1977institutionalized, salancik1978social, dimaggio1983iron, oliver1991strategic}. By behaving in ways that conform with institutional expectations, emerging organizations can improve their social and cultural fitness; however, institutional expectations sometimes conflict with each other and also with economic pressures. As a result, emerging organizations such as AI startups must skillfully navigate a complex gauntlet of social, cultural, and economic challenges. How these field-level dynamics factor into the ethical choices of startups, such as their decisions around the use of AI, is an under-explored area of research.

Here, we contend that before effective ethical AI practices for startups can be developed, an understanding of the inter-organizational and institutional dynamics these firms face must be developed. Building on recent scholarship that takes a contextual and organizational approach to ethical AI, we engage in descriptive research around the life of ”AI” as a discursive concept and organizational practice that is situated within an institutional context. Instead of focusing on ethical practices directly, we take a step back to ask fundamental questions about the forces that shape the very nature of how entrepreneurs define, build, and talk about AI itself. 

To that end, we ask two research questions:\\

\textbf{RQ1}: What institutional pressures influence how startup entrepreneurs define, discuss, and build AI?\\

\textbf{RQ2}: When do entrepreneurs comply with, avoid, or resist these pressures? \\

To address these questions, we conducted semi-structured interviews with 23 individuals working at early-stage startups across a range of industry domains. In our interviews, we focused on the financial, regulatory, and normative pressures AI startups encounter. Using abductive analysis, we illustrate how AI entrepreneurs both comply with and resist institutional pressures through the technological and business practices they employ. We find that AI entrepreneurs' face a tension between the expectations of technology entrepreneurship, which rewards rapid development and optimistic promises about technology's potential, and entrepreneurs' own values of scientific integrity, which prioritize meticulous practices and encourages skepticism. This tension was further heightened by external stakeholders' unrealistic expectations about the potential of AI, particularly when such stakeholders had limited technical knowledge. We also find that whereas AI entrepreneurs saw privacy regulation as beneficial and aligned with their own values of autonomy, they held less uniformly positive views of other AI regulatory processes such as those employed by the Food and Drug Administration (FDA) in approving AI medical devices. Drawing from our theoretical motivations, we conclude with a discussion of how our results point to both constraints and opportunities for future research on ethical interventions for AI startups.

\section{Related Works}

Organizational dynamics are a significant source of influence on the effectiveness of interventions for ethical and responsible AI. Practitioners in well-resourced organizations have expressed aspirations for ethics-supportive structures as such cross-team integration, risk-anticipation frameworks, and firm-level mission and values \cite{rakova2021responsible}. Our own research builds on these recent findings by addressing inter-organizational and institutional conditions (i.e., external pressures) that are likely to be sources of change for organizations, especially emerging organizations such as AI startups, which typically lack formal mechanisms for addressing ethical concerns \cite{vakkuri2020just, hopkins2021machine}. In this section we provide an overview of the relevant frameworks we draw from in organizational theory to explain how organizations adopt procedures and adapt over time to institutional pressures. 

\subsection{Resource Dependency and Institutionalism: How Organizations Change}

Organizational theory offers different frameworks to understand how organizations change. In the early days of the discipline, the dominant paradigm was of rationality: theorists described organizations as rational systems, machines for achieving a goal in the market, and that all organizational decisions were imbued with this same mechanical, systematic precision. Within this school were early 1900s thinkers like the German sociologist Max Weber, with his focus on bureaucracies as structural realizations of rational authority \cite{weber1978economy}, and American mechanical engineer Frederick Winslow Taylor, with his focus on bringing "scientific" methods to management to wring ever-greater "efficiency" out of a labor force \cite{taylor1919principles}. 

Starting in the 1970s, however, the field took a relational turn, recognizing that organizations do not operate in a vacuum of rationality but instead within complex ecologies of other actors. Two schools of thought--resource dependency theory and institutionalism--both address how firms seek to mitigate external pressures and uncertainty within their organizational ecosystems. Resource dependency theorists focus on organizations as their unit of analysis, i.e. the "meso" or "middle" level of institutional change (bigger than "micro", or individual people, but not as large as "macro", or field-level norms or systems). They examine the interactions between these units, using this perspective to analyze how organizations strategically seek to manage resources and mitigate dependencies on their exchange partners \cite{salancik1978social}. In doing so, organizations improve their fitness within the market. Resource dependency theory has been recently used to analyze the precarity of firms operating within complex supply chains, as they "require networks to accommodate the interdependencies in product and service flows, resource flows, and information flows" \cite{olan2022sustainable}.

The theory of institutionalism looks at the "macro" level, foregrounding patterns taking place at the level of entire organizational fields or social orders. Institutionalism focuses on unconscious social and cultural expectations, contending that these influences lead to widespread changes in multiple organizations, shaping fields of industry \cite{dimaggio1983iron, meyer1977institutionalized, zucker1977role, van2019social}. A critical component of institutional theory examines how new organizations establish legitimacy, where their actions are perceived as "desirable, proper, or appropriate within some socially constructed system of norms, values, beliefs, and definitions" \cite{suchman1995managing}. Amidst technological and market uncertainty, new firms improve their odds of survival by accruing legitimacy from an audience of stakeholders in the field including funding entities, regulatory bodies, and competitor companies. In their pursuit of legitimacy, organizations change over time, increasingly reflecting the established norms and practices of the field \cite{dimaggio1983iron}. Recent scholarship has used this lens to examine organizational changes around the implementation of novel technologies and practices such as how news publishers choose to implement novel cybersecurity tools \cite{watkins2019managing} and how AI governance and accountability (i.e. algorithmic impact assessment) may become more widespread \cite{selbst2021institutional}. 

Both frameworks have limitations in their explanatory power; resource dependency theory under-emphasizes sociocultural forces and institutionalism under-emphasizes organizations' instrumental actions \cite{scott2013institutions}. Oliver \cite{oliver1991strategic} argues that neither institutionalism nor resource dependency can adequately capture the complexity of organizational action and evolution within an institutional context. She theorizes that in addition to exercising agency in response to market demands, organizational actors can also respond strategically to institutional pressures through a variety of compliance and resistance tactics. Oliver details five core strategies organizations adopt in response to institutional pressures varying from passive compliance to active resistance. On the compliance end of the spectrum, she places "acquiescence" in which organizations conform to institutional expectations. She describes "avoidance" and "compromise" as partial compliance strategies in which organizations, for example, attempt to disguise their non-conformity with institutional norms or attempt to balance the sometimes conflicting pressures of institutional stakeholders. On the resistance end of the spectrum, she places "defiance," in which organizations actively and openly reject institutional norms, and "manipulation," in which organizations attempt to co-opt, supplant, or control institutional pressures. 

Oliver also theorizes organizations' likelihood of engaging in compliance or resistance strategies will depend on a variety of institutional conditions. The greater the perceived benefit of institutional conformity, the likelier organizations are to comply with such pressures. On the other hand, competing stakeholder expectations, multiple conflicting institutional norms, or conflicting organizational goals and institutional pressures are likelier to engender resistance. Moreover, when organizations are coerced into conformity via legal means, they are likelier to resist as compared to when institutional conformity is effected through diffuse institutional norms that are adopted voluntarily by organizations. Through the exertion of skillful agency, organizations can both gain legitimacy through selective compliance with institutional norms, while also maintaining the practices they adopt in the service of market fitness \cite{oliver1991strategic, fligstein1997social, suddaby2005rhetorical}. 

In our current work, we draw on Oliver's framework \cite{oliver1991strategic} to conceptualize how AI startups attempt to manage institutional and market pressures through a variety of compliance and resistance strategies. We consider the conditions that contribute to startups' responses to institutional and inter-organizational pressures and discuss how such strategies might interact with or constrain the ethicality of applied AI.

\section{Methods}

In this section, we describe our methods for gathering data to answer our research questions. We elected to conduct qualitative interviews, as interviews are an ideal method for better understanding actors' cognitive interpretations of their social reality, and for accessing their own explanations of their behavioral practices within that social reality. 

\subsection{Participant Recruitment \& Sampling}

We recruited participants, which we refer to here as "entrepreneurs," from US-based, early-stage startups that involve a significant AI, machine learning, or predictive analytics component. Our focus on early-stage startups was driven by theory \cite{palinkas2015purposeful}; because emerging organizations face many threats to their survival, they are heavily dependent on other organizations. As a result, inter-organizational and institutional dynamics are likely to factor significantly into their behavior. We define “early-stage” as companies that had raised less than \$50M in funding from any source \textit{or} who were at or before the Series B stage.  As our interviews progressed and patterns related to both regulatory pressures, especially privacy and the FDA, and related to funding, especially VC and crowdfunding, we began targeting our recruitment efforts towards additional startups that would further illuminate these trends. We recruited participants through a variety of methods including: 1) posts to AI and technology related listservs, message boards, social media, and Slack groups; 2) messages to general company contact email addresses or to specific individuals through email or LinkedIn; and 3) through our own direct or indirect professional contacts. In total, we interviewed 23 entrepreneurs from 20 different companies. Our entrepreneurs' companies came from a variety of industry domains including healthcare (\textit{n}=7), business analytics (\textit{n}=6), fitness and wellness (\textit{n}=5), design and engineering services (\textit{n}=2), aviation (\textit{n}=1), social planning (\textit{n}=1), and agriculture (\textit{n}=1). A breakdown of the self-described demographic, educational, and professional characteristics of our sample are available in the Supplementary Materials \footnote{https://arxiv.org/abs/2203.01157}. Participants were sent a \$25 gift certificate in exchange for their participation.

\subsection{Interview Protocol}

At the beginning of the interview session, we described to our entrepreneur participants our practices for protecting their privacy and confidentiality, then read them a verbal consent script, and gave them an opportunity to ask any questions. After providing consent, the first author asked questions based on the interview instrument (provided in the Supplementary Materials). In general the interview instrument was designed to surface data relevant to several core areas related to our research questions, including questions about the overarching aims of the company and the entrepreneur's role within it, followed by questions tailored to entrepreneur's area of expertise. For example, in cases where entrepreneurs were involved in the company's AI development, we asked additional technical questions about data collection, choice of models, evaluation criteria, and infrastructure. In most cases, we asked entrepreneurs about their companies' existing sources of financing and their plans for fundraising. We also asked entrepreneurs for their personal definition of AI. Lastly, we asked entrepreneurs about the social or ethical implications of their company. Audio recordings of the interviews were sent to a third party service for transcription, which were then verified by the first author. Our study design and protocol were approved by the Princeton University Institutional Review Board (IRB).\\

\subsection{Data Analysis}

We adopted an abductive approach to our analysis \cite{timmermans2012theory, tavory2014abductive}, which allowed us to iterate between deductive analysis guided by relevant theory and inductive analysis guided by emergent patterns in our data. To facilitate this process, the first author initially selected 11 transcripts that contained discussions of theoretically-meaningful themes or that illustrated common patterns across participants. In a preliminary analysis phase, both authors read each of these transcripts and applied descriptive codes \cite{saldana2013}. After discussing these codes in detail, transcripts were re-coded using line-by-line in-vivo codes (i.e., using participants' own words) in an effort to better preserve entrepreneurs' perspectives \cite{saldana2013}. For example, one participant discussed the drawbacks of venture capital (VC), relating VC financing to rocket fuel: "There are actually very few businesses where rocket fuel is the right thing [P13]." This excerpt was tagged with the in-vivo code "rocket fuel." In-vivo codes were then aggregated into groupings of similar topics. For example, codes related to VC funding were grouped with the "rocket fuel" quote. The authors subsequently discussed the in-vivo codes as well as relevant theory and chose to focus the next analysis phase on five core themes: 1) the "AI hype cycle" or how "buzz" surrounding AI drives external stakeholders' interest in AI companies; 2) practices surrounding the scientific legitimacy of entrepreneurs' AI approaches; 3) pressures to raise funds or secure clients; 4) the impact of regulations; or 5) entrepreneurs' own personal beliefs and ethical values that relate to their companies. These themes all appeared in multiple interviews with participants and had direct relevance to the study's theoretical focus on institutional and organizational theory. All 23 transcripts were then coded according to the five themes. The authors frequently discussed transcripts and code applications to achieve consensus. We did not measure inter-rater reliability. Inter-rater reliability is a methodologically unhelpful tool for interpretive research, when codes comes out of the collaborative process between researchers and consultation with literature, and not emergent ground-up from data \cite{mcdonald2019reliability}.

\section{Results}

\subsection{Organizational Responses to Financial Pressures}

The need to signal legitimacy to sources of financial support (e.g., investors and clients) constituted a significant vector of influence on how entrepreneurs defined, spoke about, and developed practices for AI. A tension that arose repeatedly in our interviews derived from a conflict between institutional values, specifically the values of science as a practice, and the values of technology entrepreneurship. Whereas scientific practice values systematic, methodological approaches paired with conservative interpretations of findings, technology entrepreneurship values rapid innovation and aspirational visions that extend beyond current technological reality, i.e., the "fake it 'til you make it" Silicon Valley culture. In a variety of ways, AI entrepreneurs attempted to mitigate the conflicts between the values of entrepreneurship and science by decoupling their external rhetoric from their day-to-day practices.

One way this decoupling manifested was through the use of the idea of "AI" itself. Entrepreneurs leveraged the concept of "AI" as a symbol of their technological proficiency even though they personally harbored disdain for the technical ambiguity of the concept. According to our entrepreneurs, "AI" had no precise technical meaning and was instead employed as an operational tool to signal legitimacy to resource-rich external stakeholders [P1, P2, P5, P6, P7, P9, P10, P11, P13, P14, P15, P18, P21] rather than as an accurate descriptor of what their companies actually do. In other words, "It's just a buzz word [P6]" for primarily marketing benefit. 

Entrepreneurs described a widespread belief that companies benefited from marketing themselves as "AI" companies regardless of the nature of their underlying technology. They expressed frustration with their peers who "got to use the hype term [P18]" without employing any technical practices that entrepreneurs judged as legitimate. Entrepreneurs described feeling annoyance with these AI imposters, but nevertheless admitted to employing the same marketing tactics themselves. Faced with a competitive landscape in which startups' technical and business value cannot be objectively verified, entrepreneurs leverage the institutional expectations around the legitimacy of AI because it "gives credibility that we're on the cutting edge of stuff [P2]." 

Despite the prevalence of this narrative in our interviews, only rarely were entrepreneurs able to provide specific explanations or concrete examples of how the abstract idea of AI yielded a tangible benefit. One entrepreneur, however, pinpointed investors' fear of missing out on deals as a key driver of the "AI hype cycle [P6]":

\begin{quote}
I think [the AI space] feels very confusing to [investors], but they also feel like there's every signal that it's super lucrative. [...] The key thing that keeps all of the subordinates [up at night], the ones whose job it is to go find those deals and make sure their bosses don't miss any great deals [...] is a version of the world, where you passed on Lyft. And then your boss comes back to you five years later and is like, "I would've made a billion dollars off of Lyft. [...] What's wrong with you? [P10]" 
\end{quote}

Entrepreneurs emphasized that although some investors and clients have AI expertise, most do not have the technical background required to adequately evaluate an AI solution or were simply "totally disinterested in the technical details [P6]." As a result, entrepreneurs face institutional pressures to describe their technology using homogenized, hyped language about AI even though their underlying algorithmic approaches were heterogeneous and often carefully devised. Through their instrumental use of hyped AI messaging, AI startups engage in what Oliver \cite{oliver1991strategic} refers to as "concealment." Externally startups affect the appearance of compliance with institutional expectations surrounding technology entrepreneurship even as their internal practices diverged, often significantly, from this affectation; "AI" became a discursive tool for avoiding institutional pressures via a process of "window dressing" \cite{oliver1991strategic}. 


In contrast to their external messaging, internally, entrepreneurs sought to achieve high standards of scientific rigor and validity. For example, entrepreneurs emphasized their rigorous data selection and curation processes [P4, P5, P18], described checks on the validity of their systems [P4], designed algorithm evaluations to appropriately assess the performance of their systems [P2, P4, P5, P9], and even employed independent validations by academic collaborators to ensure that their models had good generalizability [P5]. Entrepreneurs also described translating methods and findings from the academic literature in their products [P7, P17, P15] and employing scientific subject matter experts either directly on their development teams or indirectly through advisors or boards of directors [P7, P17, P18].

Yet these scientific priorities could engender serious conflicts with the priorities of users, clients, and investors. For example, when entrepreneurs attempted to use scientific legitimacy as a differentiator in pitches, this attempt was sometimes regarded by investors or potential clients as confusing or unconvincing. Two entrepreneurs described remarkably similar experiences, in which presentations about the scientific merits of their technologies were dismissed by investors as being merely "science projects [P5]" or "high school projects [P3]," having little business relevance, which from one entrepreneur's perspective, felt "like an anti-science trivialization of what scientists do [P5]."  

External stakeholders' beliefs about the potential of AI also created a barrier for entrepreneurs to be readily transparent in their external messaging about their models' methodological strengths and weaknesses. Instead, dovetailing a finding briefly touched upon by \cite{hopkins2021machine}, we observed institutional pressure around how "quality" in AI ought to be reported [P4, P5, P14, P21], specifically in external stakeholders' arbitrary notions of what constitutes "good" model accuracy. Though an algorithm's accuracy may seem objective, in practice, accuracy metrics involve many subjective choices. For example, the practical applicability of a measure of an algorithm's accuracy is contingent upon its mathematical formulation (e.g., area under the curve (AUC), F1, sensitivity, etc.) as well as contextual relevance  (e.g., the severity of a false positive versus a false negative for a medical test) and which data are selected. In our interviews, entrepreneurs felt that in order to obtain the resources necessary for their companies to survive, they needed to engage in rhetorical messaging that complied with stakeholders' expectations about model performance, even if these metrics were not the best reflection of the task at hand, nor a valid reflection of their algorithms' capabilities. 

\begin{quote}
     How it's measured is we have to make sure it's 90\% or above [...]. So if we need to switch from top 3 accuracy to top 5, just people seeing a 9, they don't even think about what it's measuring ... People just have artificial concepts of what's good and what's bad [P4].
\end{quote}

Pressure to present model metrics that have the right "psychological effect [P14]" on outside stakeholders was in conflict with entrepreneurs' desire to adopt methodologically rigorous AI approaches internally. Strikingly, one explained that his attempts to include diverse training data in the service of higher out-of-sample generalization damaged his company's credibility when his models' performance was compared to competitors who use less realistic data.    

\begin{quote}
Ultimately, our results aren't going to be as stellar as a lot of others because now we have to account for [...] all the variability within the data set whereas, if we're just focused on one homogeneous data set, our accuracy stats will be higher. So, that has been one sticky, difficult point in terms of head-to-head comparisons [P5]
\end{quote}

In the service of survival, entrepreneurs sometimes conformed at least superficially to the pressures of stakeholder expectations, adapting their external messaging over time to provide a level of scientific detail that was persuasive to the target audience. In this way, they again engage in what Oliver refers to as concealment tactics. However, these pressures did not entirely undermine their desire to externally project the methodological rigor they prioritized internally. Instead, entrepreneurs chose to target their products to specific stakeholders who would be more receptive to messages about the product's scientific credibility or its technical utility within a domain. For example, one entrepreneur highlighted that having extensive scientific references available on their product's website attracted desirable early users:

\begin{quote}
    "[...] most marketers are like, "I don't think that sells the product," but we disagree. [...] I'm not sure it makes it so it's a blockbuster of a product, but it brings in the right type of people for your product [...]. It brings in good early adopters, anyways [P7]"
\end{quote}

In this public commitment to scientific integrity, entrepreneurs engage in a form of "defiance" \cite{oliver1991strategic}. Oliver hypothesizes that defiance is more likely when the perceived cost of resistance to institutional norms is low and "when they can demonstrate the rationality or righteousness of their own alternative convictions or conduct." AI entrepreneurs who externally project their strongly-held personal values of the scientific process may do so because they can promote science as a virtue while still attracting science-inclined external stakeholders.

Entrepreneurs recognized that in order for their companies to grow, their products must eventually translate into market success and that their companies' investors were ultimately motivated by whether or not the company would "provide liquidity [P17]" on their investment. Nevertheless, entrepreneurs demonstrated a wide range of compliance and resistance tactics when it came to their decisions regarding financial backing. They described pursuing a variety of strategies to fund their businesses including revenue [P8, P13, P21], friends and family raises [P1, P10], grants from government or private entities [P2, P12, P17], angel investors [P4, P5, P6, P7, P13, P14, P15, P17, P21], debt financing [P12], VC financing [P4, P10, P14, P16, P17, P18, P20], and crowdfunding [P7, P9, P13, P21, P23]. Even still, several entrepreneurs pointed out that VC is regarded as a default financing path:

\begin{quote}
    We looked at the VC funding route in the beginning because that's what you're told to do, right? That's how you get funded. This is the path. You go pre-seed, it's angel investors, after that it's VCs, and then you go through the Series process [P12].
\end{quote}

For some, capital from VC firms formed a cornerstone of their strategy for building their business [P4, P10, P16, P17,P18, P20]. These entrepreneurs saw investors not only as a source of financial capital, but also of valuable industry domain and business expertise as well as a mechanism for accessing important professional networks. Entrepreneurs viewed the fit between the needs of the company and the expertise of investors as a critical component of establishing a productive relationship. 

\begin{quote}
    So, we have a number of investors, and the asks really change based on business needs. So it's really, what do we really need today, this week, this month, that can help us take the business to the next level, and who do we have as investors that we can ask for help in those areas? [P17].
\end{quote}

However, entrepreneurs did not blindly acquiesce to the demands of investors. Where their own goals conflicted with the goals of investors, VC or otherwise, they would sometimes decline further involvement. For example, one entrepreneur described ending early conversations with an investor because their desired exit strategy was not consistent with her own goal to eventually take the company public [P1]. In another case, an entrepreneur described evaluating potential investors based on their alignment with the company's ethical values: 

\begin{quote}
    We are trying to raise capital from investors that have the same kind of values and mindset with us and people who are not afraid to lose certain revenue or sales just to follow the same values. We had clients asking us to do things that we said, you know what? No. No, this is not something we feel comfortable with doing [P23].
\end{quote}

One entrepreneur noted that because of intense investor interest in the field of AI, instead of entrepreneurs doggedly pursuing financiers, "they find you [P10]." In a resource-rich environment, entrepreneurs have more latitude to resist institutional pressures arising from dependencies on investors. To some extent, we saw this resistance in how entrepreneurs described choosing specific investors; however, we saw even greater resistance amongst entrepreneurs  who expressed hesitancy about pursuing VC financing [P7, P9, P12, P13, P16, P21]. In these cases, VCs were seen as having financial goals that conflicted with entrepreneurs' long-term business or product objectives. One entrepreneur noted that unlike traditional software, AI software development typically requires specialized algorithm expertise, infrastructure, expensive data labeling, and continuous model performance monitoring; however, if VC firms' valuation of AI companies is based on their knowledge of traditional software startups, they may impose timelines or key performance indicators that undermine what entrepreneurs believe to be methodologically-sound practices in AI development. Recognizing that VCs' profit motives and responsibility to their limited partners (LPs) could conflict with their own goals, entrepreneurs often described actively avoiding VC financing.

\begin{quote}
    That was my whole experience [at a previous startup]. The VCs wanted to hype things up, get a lot of press, make a splash, so they could raise the next round at a higher valuation and look good to their LPs, which was actually contrary to what we needed to do for the slow growth to build the business [P13]
\end{quote}

Instead of VC funding, several entrepreneurs described using crowdfunding to finance their businesses [P7, P9, P13, P21, P23]. Crowdfunding and other financing vehicles that provide investors with little direct control over the companies' behavior were viewed by entrepreneurs as a way to maintain autonomy over their businesses, control progress towards the product vision, and to maintain the equity of current employees. Moreover, although prior literature has indicated that the disclosure requirements of crowdfunding platforms can create a risk for companies' subsequent financing prospects \cite{blaseg2021equity}, entrepreneurs valued the transparency associated with public disclosures on crowdfunding platforms and the opportunity to directly engage with potential crowd investors. The choice of crowdfunding over the "default" VC financing path constitutes another form of organizational defiance; these entrepreneurs challenge the culturally dominant mode of startup funding by choosing financing paths that they felt would better serve their long-term objectives. As predicted by Oliver \cite{oliver1991strategic}, startups are able to engage in resistance because doing so does not compromise their chances of survival since they can rely on alternatives to VC to fund their businesses. 
\subsection{Organizational Responses to Regulatory Pressures}

In addition to financial pressures, entrepreneurs also described their compliance with and resistance to pressures in the form of regulation, especially regulations surrounding privacy and the Food and Drug Administration (FDA) approval processes for applications of AI in medicine. On the whole, entrepreneurs viewed privacy protections as normatively good, a competitive necessity, or even a competitive advantage compared to industry peers who are slower to adopt privacy-protecting practices. One entrepreneur described how his company had chosen to temporarily avoid adapting to the "landscape of privacy and privacy laws [that are getting] a lot more strict [P19]" by operating selectively in markets that were subject to less stringent privacy regulation (e.g., in the US versus European Union where GDPR is applicable), but even this entrepreneur noted that circumvention was only tenable in the short term. Yet in contrast to \cite{hopkins2021machine}, most entrepreneurs in our study did not express resistance to or subversion of privacy regulations but openly endorsed them as well as the ethical values underlying them, such as personal autonomy. In their discussions about privacy, entrepreneurs sometimes contrasted their own beliefs and policies with those of large technologies such as Facebook, whose privacy-related behaviors they generally regarded as reprehensible [P9, P6, P23].

Unlike privacy regulation, the FDA approval processes for AI in healthcare was viewed less favorably. Entrepreneurs who discussed the FDA viewed these regulatory requirements as unnecessarily onerous and in some cases, unscientific [P2, P4, P5, P7, P16]. A theme that arose in our interviews with multiple entrepreneurs was that the lack of standardization in the FDA approval process for AI-enabled healthcare products created a serious burden for startups with an unclear upside. One entrepreneur from a more mature startup described how challenges posed by the FDA approval process contributed to her company's decision to eventually pivot away from the AI products she felt had life-saving potential to a core product and business model that she thought was more profitable but ultimately less useful to society. She highlighted the opportunity and financial costs of pursuing FDA approval as well as the difficulty of providing evidence to meet FDA performance standards: 

\begin{quote}
 Whenever you're thinking about a rule-out device–in our case, [the finding that a medical test is normal]--that means you rule out every possible thing, it’s statistically insanely hard to do. And in order to get approved, you would literally have to do better than humanly possible [P16].   
\end{quote}

All but one entrepreneur [P18] who discussed relevant FDA regulation found the ad-hoc approval process to be legally or financially arduous; however, there was less consensus on whether the FDA's model performance standards were unreasonable. One entrepreneur noted: 
\begin{quote}
    The data that needs to be provided in order to get clearance in our opinion is relatively low, but it does take a lot of money and other things to get [P5]
\end{quote}

Similar to this perspective, another noted that if the financial demands for securing FDA approval were lower, they would begin reallocating their research and development efforts to undergo FDA approval as soon as possible since their model performance was already strong. 

Entrepreneurs either implied or stated explicitly that the onerous FDA approval process stifled innovation, but they also noted that such regulations were necessary to protect users from harmful products. Another entrepreneur noted that regulatory approval could even be beneficial to his businesses because "customers are much more receptive to the FDA stamp than they are to stats [P5]". Still, the perception that the FDA approval process was "a whole monster [P4]" that was often not worth pursuing motivated AI entrepreneurs to attempt to operate in regulatory gray areas or exploit loopholes so that they could continue to pursue technological or product objectives. Such entrepreneurs described using other, non-regulatory avenues to demonstrate their legitimacy such as publishing their model details and performance in academic journals or technical whitepapers. It is important to note that entrepreneurs' opposition to the idiosyncratic FDA approval process was not merely a matter of logistical difficulty; they also viewed the discretionary nature of the FDA review process as an opportunity for established companies to be unfairly advantaged:

\begin{quote}
    It's also, I think, unfair how the FDA, [...] they have existing relationships with Pfizer and Johnson \& Johnson. I get it. But they're obnoxiously hard on startups because they're not known [P16]
\end{quote}

Thus, entrepreneurs' personal values only partially aligned with the institutional pressures from the FDA--they value the consumer protection intent, but decry its consequences for innovation and believe it to be at odds with a fair, competitive marketplace. In contrast, entrepreneurs were more likely to adopt privacy preserving practices, which were consistent with their own normative beliefs, even in cases when companies were not yet subject to strict privacy regulation. In other words, consistent with Oliver's \cite{oliver1991strategic} prediction about legal coercion engendering less compliance than institutionally diffuse norms, there appears to be an association between internal adoption of institutional rules derived from regulation when such rules aligned with field-level values.  

\subsection {Organizational Responses to Technological Pressures}

Institutional theorists posit that emerging organizations can bolster their own legitimacy by adopting the values, structures, and practices of established organizations \cite{meyer1977institutionalized, dimaggio1983iron}. That is, organizations can improve their chances of survival by mimicking what incumbent organizations already do. Yet this idea is in tension with the Silicon Valley notion that the most legitimate innovations are those that "disrupt" existing ways of operating \cite{hogarth2017valley, geiger2020silicon}. So-called radical innovations are those that break from or are discontinuous with prior scientific and engineering practices whereas incremental innovations are those that build upon and extend the existing technological paradigm \cite{colombo2015going}. In our interviews with entrepreneurs, their rhetoric typically suggested they viewed deep learning models as constitutive of radical innovation in that they distinguished deep learning models from other machine learning techniques; however, their implementations of deep learning models--typically via transfer learning--were, in contrast, fundamentally incremental.

Despite the murkiness around definitions of AI more broadly, entrepreneurs often held deep learning out as distinct, which sometimes manifested in how entrepreneurs defined AI. Many entrepreneurs provided definitions that contrasted algorithms' capabilities with human capabilities [P6, P7, P14, P16], that differentiated between general and narrow intelligence [P5, P6, P7, P15, P23], or that described high-level processes that are applicable to any AI model [P1, P2, P4,P7, P8, P15, P18, P19, P20, P21, P22, P17]. However, several entrepreneurs implicated deep learning specifically in their definitions [P2, P9, P13, P11], using deep learning as a threshold of "real" AI: 

\begin{quote}
    The most concise answer I can give you is just deep learning. That is almost the new cutoff for AI, in my mind at least [P9]. 
\end{quote}

Even for those that did not equate deep learning with AI, the ways entrepreneurs discussed deep learning relative to other AI approaches suggest that they consider such techniques separate or superior. For example, some described using "machine learning and deep learning [P14]" or explained that their companies constrained their models to "classifiers and regression [P3]" instead of "doing any deep learning or anything nutty like that [P11]" as though deep learning models were not a subset of machine learning techniques. In one case, we interviewed the chief technology officer of a company that intended to develop an AI-enabled solution but that had not yet begun data collection. Even in the absence of any empirical evidence to support his conclusion or expertise in deep learning, he had preemptively concluded that simple, linear techniques would be insufficient to achieve the high accuracy he hoped to obtain with deep learning methods [P14]. Thus, entrepreneurs distinguish the "magic [P13]" of deep learning from other "rudimentary data science [P11]" techniques.

Yet in conflict with widespread framing of deep learning as "magic" and perhaps radical innovation, most entrepreneurs' implementations of deep learning constituted a more incremental form of development that draws on the scientific products of researchers in the AI community. Many entrepreneurs described relying heavily on transfer learning for their deep learning applications. Transfer learning is technique where models that are initially trained with massive datasets for one task can be adapted for related tasks with much lower data requirements. The use of transfer learning can reduce the computational and data costs associated with training models from scratch while still affording entrepreneurs "the accuracy that we feel we need from that model [P9]". Through the use of pretrained models initially developed by AI researchers in academia or at large technology companies, AI startups "build upon the state of the art, all the advancements that are being driven by the Googles of the world [P13]". In a paradoxical way, entrepreneurs' discursive distinction of deep learning techniques complies with the institutional pressures to seek rapid and disruptive innovation, even though the pretrained models used for transfer learning coordinate practices of AI entrepreneurs, possibly down to the specific pretrained models they employ. Entrepreneurs' rhetoric surrounding the distinction of deep learning, which conforms with institutional expectations about the utility of disruptive technology, acts to conceal their use of publicly available, incremental technologies that while not fundamentally disruptive in a scientific or technical sense, are nevertheless sufficient to meet AI startups' needs. Interestingly, the disconnect between entrepreneurs' rhetoric and practices in deep learning did not appear to be driven by attempts to appeal to external stakeholders who would be unlikely to appreciate the difference between deep learning and other machine learning techniques, but potentially to signal status to other AI startups or industry peers or to bolster their sense of the company's legitimacy internally.



\subsection{Organizational Responses to Normative Pressures}

As an organizational field becomes more institutionalized, professionalization is enacted through education, membership in professional bodies, and other aspects of professional culture. These professional mechanisms can drive organizations within that field to adopt similar norms and values, which become embedded in their organizational practices \cite{dimaggio1983iron}. Recent scholarship on professional norms within the AI research community has found that pervasive professional norms include efficiency, universality, and impartiality \cite{scheuerman2021datasets}. In our interviews, we observed instances where entrepreneurs articulated personal ethical values that were either distinct from, or resistant to, professional norms.

The demands of operating within the fast moving technology industry constrain the extent to which industry practitioners can fully realize ethical values into substantive practices \cite{vakkuri2020just, metcalf2019owning}. Consistent with these findings, some entrepreneurs hoped to incorporate their personal ethical values into their product or business model in the future, but had yet to make much tangible progress towards those ideals [P1, P2, P10, P13]. Yet, in other cases, entrepreneurs took a strong stance on ethical issues and described how they built these values into their technology and organizational cultures. For example, several entrepreneurs described how their algorithms [P21, P23, P19, P6] or data practices arose from normative beliefs about the ethics of privacy protection [P2, P9, P11, P14]. 

Racial bias also came up repeatedly in our interviews, but did not always inform product or business decisions. In a handful of cases, startups' AI approaches had been explicitly designed to ensure that their algorithms would perform equally well across demographic groups [P4, P21, P23]. Similarly, some entrepreneurs had designed non-algorithmic elements of their products to prevent racial bias [P17, P12]. In several cases, entrepreneurs' motivations for developing algorithms that perform well across racial groups were not only based on personal value systems, but also based on the belief that fair algorithms realized market value. For example, several entrepreneurs pointed out that in order to serve international clients and diverse users, it was important for AI-enabled products to be equitable. On the other hand, some entrepreneurs were aware of the types of racial biases that can be reproduced by AI algorithms [P6, P7, P9, P11, P22], but either thought that race was irrelevant to their models [P6, P11] or that racial biases were only a priority in high-stakes contexts such as healthcare and finance [P9, P11].

Although less common than algorithmic strategies, several entrepreneurs touched on how they promoted racial and gender equity within their companies [P12, P22]. Drawing from his own experience of racial marginalization, one entrepreneur noted:

\begin{quote}
    A fundamental shift in power from straight white men to the rest of the world is really something that needs to happen. [...] I want to be able to show people that look like me, that they can also use things, and they can also build something that's great and can also help build those communities [P22].
\end{quote}

Even outside of explicit interview questions about ethics or social impact, entrepreneurs often espoused values related to democratization and expanded access to technology [P1, P8, P17, P16, P19, P4, P2] (e.g., "democratizing access to data [P17]"). Entrepreneurs described wanting to provide financially valuable expertise or insights to other businesses, especially small businesses and startups [P1, P19], to provide needed services to emerging economies [P16], or to empower users to take on tasks that are more typically performed by specialized professionals [P4, P8, P16]. In line with these values, entrepreneurs were critical of insider cultures, implicating "old boys'" networks [P6, P7, P22] or "traditional male VC [P12]" in gatekeeping behaviors related to client acquisition, external financing, or in other ways that affected their businesses' success. Entrepreneurs' skepticism of centralization and insider culture was also manifested through their choices about funding. Whereas VC served as a stand-in for centralized power, crowdfunding was viewed as consistent with the ideal of democratization since crowd investors do not need to meet the same financial accreditation standards required to invest in a VC fund.

Entrepreneurs' normative values reflect a mix of the techno-libertarian leanings of Silicon Valley that have been documented elsewhere \cite{metcalf2019owning, lenhard2021, dahlberg2010cyber, hutten2019soft} as well as beliefs in social equity and fairness. Sometimes these values conflicted, as is highlighted by the tension entrepreneurs expressed around racial bias in AI; they believe that all users should be treated equally, but under the same resource-constrained system that encourages developers to "move fast and break things" \cite{vardi2018move}, they do not always prioritize development around that belief.

\section{Discussion}

Our current study adds to the growing literature on organizational challenges to ethical AI by describing how broader inter-organizational and institutional forces shape the practices of AI startups. In this section, we discuss both the theory-based as well as pragmatic contributions of our research. This discussion is structured along the same categorical lines of our findings, discussing in turn financial, regulatory, technological, and normative pressures. 

\subsection{Financial Pressures}
 A central tension recurred between entrepreneurs’ desire to preserve the scientific integrity of their AI approaches and the demands of technology entrepreneurship that often ran counter to this desire. As one entrepreneur noted, "the value in the technology that you use doesn't necessarily even have to come from the technology [P9]". In contrast to purely scientific enterprises, the import and meaning of novel technologies is not entirely determined by scientific inventiveness or rigor, but is also constructed within an economic, social, and cultural context. 

The demands of external stakeholders with power to affect the financial outcomes of AI startups exerted influence over the narratives entrepreneurs constructed about the benefits of their technology.  In response to stakeholders’ expectations of “silver bullet [P21]" AI solutions, entrepreneurs tended to adapt their external messaging accordingly, but they did not necessarily alter their internal practices. In this way, entrepreneurs engaged in a resistance strategy of concealment \cite{oliver1991strategic}, decoupling the symbolic and homogeneous marketing tactics they adopted to accrue legitimacy from business partners, from the substantive and often heterogeneous approaches they employed internally.  

That entrepreneurs placed a strong value on scientific integrity points towards an ethical opportunity within the startup ecosystem. As several entrepreneurs themselves pointed out, models with inequitable outcomes are necessarily less valid since they do not generalize well. Moreover, they are less able to realize business value since they cannot meet the expectations of diverse clientele. Thus, entrepreneurs' values of scientific legitimacy might act as a "value lever" \cite{shilton2013values} through which principles of AI ethics can be imported into AI startups. On the other hand, external stakeholders' tendency to treat decontextualized accuracy metrics as a superficial indicator for AI quality is suggestive of a risk for institutionalization of AI ethical ideals. For example, the “80\% rule” for establishing disparate impact, which has often been imported into AI fairness research without regard for its original legal nuance, may have already created an artificial standard within the research community \cite{watkins2022four}. This metric as a target could create further ethical risk if stakeholders in the AI startup ecosystem also adopt it without considering its relevance and caveats within context. Our observations around the use of "AI" as a marketing "buzz word," reflect recent concerns around "AI as snake oil" \cite{fakeaisnakeoil}, and the exploitation of AI's vague definition as a loose umbrella term. Thus, strategies that ensure that AI ethics constitute more than an ethical Potemkin facade are needed \cite{cihon2021ai}. 

Consistent with prior literature \cite{willoughby2008entrepreneurial}, entrepreneurs also demonstrated more heterogeneity in financing strategies than the culturally dominant VC-startup narrative would suggest. While some entrepreneurs conformed with institutional pressures to pursue VC funding and found benefits beyond financial capital in their partnerships with VCs, others actively avoided VC funding. This opposition was sometimes based on philosophical opposition to VC as antithetical to democratic ideals and other times informed by entrepreneurs’ personal experiences of VCs driving startups away from sound technological and business practices. Some evidence suggests that entrepreneurs' avoidance of VCs could harm their companies' growth potential \cite{baum2004picking, bertoni2011venture}, but other evidence shows that the benefits of VC do not always extend to profitability
\cite{rosenbusch2013does}. Moreover, even if VC does improve the financial outcomes of companies on the whole, this financial benefit does not necessarily redound to founders themselves since their stake in the company is significantly diluted by VC investment \cite{florin2005venture}. Thus, resistance to the institutional norm of VC financing could be conceptualized as economically rational as well as in line with entrepreneurs' desire to retain control over their businesses since VCs sometimes use their power to replace the founding team with professional executives \cite{hellmann2002venture}. 

Entrepreneurs' desire to match their financing strategies with their business goals and normative values presents an opportunity for ethical practices. Even if AI ethics interventions are seen as antithetical to profit goals as demonstrated in \cite{von2021cost}, entrepreneurs may be able to preserve AI ethical ideas by matching with investors who share these priorities, especially if public scrutiny around the ethical implications of investor strategies increases \cite{amnesty2021}. It is important to note, however, that the entrepreneurs who are able to exercise more discretion in terms of when they seek funding and from whom they seek it are likely already advantaged in the entrepreneurial ecosystem, as one of our entrepreneurs himself noted: "I know we have the luxury that we could decline money. I know that that is a luxury [P6]". Black and Latinx founders \cite{crunchbase2020} and female founders \cite{teare2021} secure less financing than other founders, and as a result are likely to have fewer options when attempting to find financing partners who prioritize ethical objectives. Thus, selective matching between entrepreneurs and investors could also further magnify inequality. 

\subsection{Regulatory Pressures}
With respect to regulatory pressures, entrepreneurs typically endorsed privacy regulations but expressed more frustration with FDA regulations. While privacy regulations were perceived as aligning with the values of personal freedom and autonomy, which have been documented in other research on technology sector actors \cite{metcalf2019owning, lenhard2021}, FDA regulations were seen as a barrier to innovation and entrepreneurial autonomy and a mechanism through which industry insiders receive favor from other institutional actors. These contrasting results support both theory and evidence that a mismatch between an organizational field's normative values and coercive regulatory pressures will result in less meaningful internalization of policies \cite{oliver1991strategic, scott2013institutions}.  However, an alternative resource-based explanation is also possible. Privacy regulations are likely to apply uniformly, but FDA regulations are idiosyncratic, and therefore require more expertise and financial resources to navigate. Regardless of the cause, as legislators debate proposals to further regulate AI, they should take care to consider what negative, second-order effects regulations might have on AI startups. Greater engagement with AI entrepreneurs could improve both policy and its adoption within startups since active participation from business owners has been shown to increase regulatory compliance \cite{malesky2017danger}. 

\subsection{Technical Pressures} 

 The resource constraints of startups also fed into our findings regarding the use of deep learning amongst AI startups. Both entrepreneurs who did use deep learning and those who did not tended to discuss the use of deep learning techniques with a reverence not afforded to other algorithmic approaches. Yet deep learning startups most often developed their technology on top of preexisting pretrained models, especially those developed to perform natural language processing and computer vision tasks. As with most scientific advancements, applications developed through transfer learning are incremental innovations, inextricably tied to established approaches developed by a broader community of researchers and practitioners. That is, the use of deep learning in most AI startups is not a radical departure from the dominant machine learning paradigm, but an endorsement of it.
 This is not to say that deep learning applications developed through transfer learning are not valuable, creative, or innovative. Experts have implicated transfer learning specifically in the acceleration of AI discovery \cite{ng2016nuts}. 
 
 
The widespread use of pretrained models does raise ethical questions. Word and image embeddings derived from models trained on human data often encode human-like biases such as gender, racial, and other harmful stereotypes \cite{bolukbasi2016man, steed2021image}. How entrepreneurs adapt pretrained models to their applications may obviate transmission of harmful biases from pretrained models to industry applications; however, some researchers have suggested that the biases of pretrained models, if not mitigated for their contextual application, could further propagate harms \cite{steed2021image}. Thus, our finding supports the call for research to better understand not only the negative social impacts of models developed in academic contexts, but also how these impacts are attenuated or magnified by their applications in industry through transfer learning \cite{narayanan2021ethics}.
 
 \subsection{Normative Pressures} As we have already discussed, entrepreneurs' beliefs played a significant role in how they developed their technologies and their business practices. In some cases, AI entrepreneurs espoused libertarian ideals, such as individual autonomy and personal responsibility. Much of AI ethics research focuses on establishing the fairness of model outputs or mitigating unfairness in model predictions. In this way, AI ethics interventions often center on equity in outcomes. In contrast, entrepreneurs expressed valuing democratization, which emphasizes the importance of equality of access, rather than equity in outcomes. This distinction points to further risks for translating the technology and ideas developed in AI ethics research contexts into AI startups or the technology industry on the whole. Institutional pressures deriving from the technology industry will likely interact or conflict with the values embedded in ethical AI interventions. Designers of ethical interventions should consider the normative context in which they are intended to apply. Otherwise, they could be dismissed as irrelevant by practitioners or be employed in ways other than how they were designed, which itself constitutes an ethical risk. 

\section{Future Research} 

Our findings on the significant influence exerted by institutional pressures on AI startups, and the variance in entrepreneurs' decision-making around compliance, avoidance, and resistance, open a number of potential research pathways. First, as mentioned above, more research is needed to better understand how the social impacts of AI models may be exacerbated through transfer learning in industry settings. Second, identifying and interviewing other stakeholders in this sector would allow us to analyze the interactive dimension of these field-level dynamics, yielding data about how investors, regulators, competitors, and customers participate in and contribute to complex system dynamics of institutionalism in AI. Findings around the alignment between regulatory pressure and normative pressure, further, suggest that such alignments lead to better take-up within organizations, and so collaboration with policy researchers could lead to the design of AI policy better positioned to act as an effective guardrail against the harms of such systems.    

\section{Limitations}

Our study design presents several limitations which may influence our findings. First, a limitation of the interview instrument was its exploratory nature. Due to the broad scope of our research questions, themes could not be identified prior to the study, but rather were identified in our data as a set of findings. As a result, we were unable to reach depth within particular themes, nor did we find that we reached theoretical saturation for any thematic category. Instead, the exploratory nature of this study identifies pathways for future research opportunities. Second, our recruitment and sampling strategies also present limitations. Our sample size was relatively small, and cannot -- and is not intended to be -- generalizable to the larger population of AI startup entrepreneurs. Within qualitative research, sample size requirements are a subject of debate, and are always a reconciliation between research interests and goals, access to participants, and maintaining rigor. We used theoretical sampling, which is intended to ensure that there are enough participants to surface "a range of concepts and characteristics that are deemed critical for emergent findings," \cite{dworkin2012sample,glaser1967discovery}, which we determined was achieved with our sample. 

\section{Conclusion}

On the whole, our research shows that although institutional forces do shape AI startups' beliefs and practices surrounding AI, they do not dictate them entirely. As a result, while future interventions to support ethical AI should be mindful of the organizational contexts for which they are intended, they also should not assume that startup practitioners have no agency to act in the service of ethical values. Even if the ethical practices adopted by startups at their outset evolve over time in response to shifting market demands, founders typically have a lasting influence on startups' trajectories, even after they leave the company \cite{sahaym2016parent}. As a result, though startups face more resource constraints than the more mature companies that have been the focus on most applied AI ethics research, they also may be an ideal stage for ethical interventions.

\section{Acknowledgments}
We thank Ranjit Singh, Arvind Narayanan, and Alex Hanna for helpful feedback on our manuscript and Pedro Gomes for guidance on our research questions. We gratefully acknowledge financial support from the Schmidt DataX Fund at Princeton University made possible through a major gift from the Schmidt Futures Foundation.
\section{Bibliography}
\bibliography{bibliography}

\appendix


\section{Interview Instrument}
The interview instrument we used to loosely structure our interview is below. We note that the primary questions we asked participants from the ideals and values section of our instrument were about social and ethical implications. Typically, we asked follow up questions based on participants' responses to this main question rather than other questions in this instrument. 

\textbf{Background}

Can you tell me a little bit about your company? 

[if not already answered] 
What is the problem your company is trying to solve?

What is your role in the company?

Can you walk me through what you did at work on a specific day recently?

\textbf{Product \& AI}

How do you define AI? 

How does your company use/want to use machine learning, artificial intelligence, or predictive analytics?

Why did you/your team decide ML/AI was the right approach?

What was your/your team’s experience in AI/ML before starting this company?

\textbf{Funding} 

How is your business funded? 

Can you describe your experience trying to secure funding? 

[If funded] 
    
How do you typically interact with your funders? What happens in these interactions? 

Have you discussed how your company uses AI with your funders?  

What do you think your funders think about AI? 

[If not funded] 
    
    Once you do secure funding, how do you anticipate you will interact with your funders?

    Have you discussed how your company uses AI with potential funders? 

What do you think potential funders think about AI? 

What is your company’s exit strategy?

\textbf{Ideals \& Values}

What would you say are the core values of your company? 

What do you think differentiates a successful startup from an unsuccessful one? 

Are there other companies that you think are good examples for your own company to follow? 

What role does AI/machine learning play in the technology industry as a whole? 

Has your team ever discussed the ethical or social implications of the AI you use in your product? 

\textbf{Demographics \& Background}

What is your title at your company?

What is your gender?

What is your race? 

What is your age?

What is your educational background?

\section{Participant Characteristics}
Participants' self-described demographic characteristics and company roles are listed in Table \ref{tab:demographics}. Note that some participants listed more than one race, and many listed more than one role. The methods through which participants were recruited are listed in \ref{tab:recruitment}. The breakdown of participants by industry are available in Table \ref{tab:industry}
\begin{table}[]
\begin{tabular}{|l|l|}
\hline
                             & \textit{n} \\ \hline
\textbf{Race}                &            \\ \hline
White/Caucasian              & 14         \\ \hline
Asian                        & 2          \\ \hline
South Asian/Indian           & 5          \\ \hline
Black/African American       & 2          \\ \hline
Middle Eastern               & 1          \\ \hline
                             &            \\ \hline
\textbf{Gender}              &            \\ \hline
Male                         & 17         \\ \hline
Female                       & 6          \\ \hline
                             &            \\ \hline
\textbf{Education (Highest)} &            \\ \hline
High School                  & 1          \\ \hline
Bachelor's                   & 7          \\ \hline
Master's                     & 9          \\ \hline
PhD                          & 4          \\ \hline
MD (or equivalent)           & 3          \\ \hline
                             &            \\ \hline
\textbf{Role} &                           \\ \hline
C-Suite (e.g., CEO, CTO)    & 11            \\ \hline
Founder/Co-Founder           & 15             \\ \hline
Division Head  & 3             \\ \hline
Other  & 4             \\ \hline

\end{tabular}
\caption{\label{tab:demographics}Participant Demographics}
\end{table}

\begin{table}[]
\begin{tabular}{|l|l|}
\hline
                             & \textit{n} \\ \hline
1st degree contact of authors             & 1        \\ \hline
2nd degree contact of authors              & 5          \\ \hline
Slack groups          & 4          \\ \hline
In-person networking event       & 4          \\ \hline
University alumni message board              & 2          \\ \hline
Cold contact             & 7          \\ \hline
\end{tabular}
\caption{\label{tab:recruitment}Recruitment Methods}
\end{table}

\begin{table}[]
\begin{tabular}{|l|l|}
\hline
                             & \textit{n} \\ \hline
Healthcare             & 7        \\ \hline
Aviation              & 1          \\ \hline
Fitness \& Wellness          & 5          \\ \hline
Business Intelligence \& Analytics       & 6          \\ \hline
Social Planning              & 1          \\ \hline
Design \& Engineering Services             & 2          \\ \hline
Agriculture            & 1          \\ \hline
\end{tabular}
\caption{\label{tab:industry}Participant Industry}
\end{table}

\section{Supporting Quotes}

Scholars have hotly debated whether the goals of open science that have been gaining traction in the quantitative sciences are also relevant for qualitative research \cite{pratt2020editorial, reinhart2016reproducibility, kapiszewski2021transparency, seale1999quality, rhoads2020whales, murphy2021ethnography}. Unlike quantitative research, qualitative human subjects often participate in research on the condition of anonymity, which precludes complete transparency. Moreover, qualitative work is often premised on the idea that the interpretation lens the researcher brings to research is itself a valuable component of any qualitative scientific pursuit. Here, we attempt to create a balance between these values by offering our own interpretation of our findings in the body of the article and offering as much transparency as possible without compromising our participants' anonymity through supporting quotes. We omit quotes from the supplement that that directly or indirectly could identify participants. To reduce the possibility that participants could be identified by patterns across their quotes, we do not provide a participant identifier for each quote, but we include the participants whose quotes are listed within each section. To align with the organization of the main article, quotes are organized according to financial pressures, regulatory pressures, and normative pressures and are in no particular order within sections.

\subsubsection{Financial Pressures}
Quotes are derived from P1, P2, P5, P6, P7, P8, P9, P10, P11, P12, P13, P14, P15, P17, P18, P20, P21.
\begin{itemize}

    \item {I think it's because there's a little bit of a sense of AI being a like magical silver bullet type solution. AI is just like this loosely defined thing that if you give to somebody, it could potentially make them more money or give you better insights or something like that, that from a more public perspective, as far as a company saying that they are an AI based company providing a service that may not use AI at all as better investment and also people, it shows or it signals that the solution could be more scalable than it is in its current fashion."}

    \item{We're going to talk about our machine learning algorithms, because from a marketing standpoint, it connotes this next generation high tech, God, it has to be good.}

    \item{And I think part of the biggest issue, and this may not be unspoken, is that everyone and their brother wants to have AI in their product, especially in healthcare right now, because it's the buzzword du jour. So basically, if you're in the physical space, you want two things: you want to say your product has AI and you want to say your product has a robot.}

    \item{Because [AI is] the sexy thing for investors. And quite frankly, coming from an engineering perspective of what I know of AI and what I know of machine learning, I actually think a lot of it is overblown and a lot of things that are called AI is not actually AI. It's actually machine learning or a learning algorithm that is kind of tweaked and people are bringing up AI just to say they have it.}
 
    \item{Just including a small amount of AI gives you a marketing edge.}
 
    \item{Plus it's a feedback loop, I guess. That's what you see in industry, everyone doing this. And getting good results by saying that, "Look, we use AI." It almost seems like, "Why wouldn't you?" I guess, whether or not you have it. While I did mention it as a pet peeve of mine it's also, I guess, understandable. Especially for people in startups, is a very competitive space. So you're trying to get every edge you can.} 

    \item{It is similar thing as saying "blockchain." You're well aware, but it's a super common thing with startups, that startups are doing. Just trying to catch people's attention with the hottest new tech.}
    
    \item{When I was doing research, it was just super focused on like, "Here's this new thing that I did that's novel. And it's state of the art and it gets 0.001\% better accuracy than this other guy's thing. It's not reproducible, but it's AI, and it's really cool." So I don't know. The field is super legitimate. How do I say this? You end up getting a lot of people who are just trying to ride that wave of legitimacy and not contribute anything substantial.}
    
    \item{I think the basics of it is -- a large dataset, train a machine learning model, you can predict many things. I think that's permeated a lot of just public understanding, scientific popular science. I think that basic equation resonates with people, even if they don't really care to understand okay exactly how does a neural network work or what is an LSTM or something like that. They might not care about those specifics but they probably see the results in their daily lives. I think there's been a lot of remarkable changes in the last 5 or 10 years with real products that people are using AI, improving them. And so, I think they can appreciate its ability, while not necessarily caring to delve too deep on the technical side.} 
 
    \item{But AI, machine learning, they're buzzwords, they make us sound smart, like we've got access to something because to them, they don't know any of this stuff [...]  INTERVIEWER: Who is that a buzzword for? Who's impressed by the word AI or machine learning? PARTICIPANT: It just puts us in a box of, and I'm out here in Silicon Valley -- so it's the nerds, the data people, they're in touch with tech, they associate us with the tech industry}

    \item{When it comes to funders, I think AI is a buzzword that everybody likes hearing on the VC side, big data, machine learning, those types of things. But I think from a marketing perspective, we're marketing to work with collaborators and hospitals or potential customers, it's really focusing on the clinical impact, that's the most important.}

    \item{At least, to me it's pretty obvious that machine learning has transformed a lot of different tech sectors. The larger scale ones certainly, from Google to Facebook. But down to even more specialized sectors. So I think there's this... I think the feeling is that AI has the capability of transforming or paradigm shifting different sectors, and if you get in on it early you can be part of that wave. And I think it's certainly been a very successful methodology in many fields across many domains. So it has a lot of demonstrated success.} 

    \item{[...] We have these hype cycles for AI throughout history of at least the last 80 years, 70, 80 years. And if you look back into the history of what we now understand as artificial intelligence, we have these just incredible claims what will be possible tomorrow or at least next year. And then we had these AI winters and all of that again. And I think sometimes in the last years, the last 10, 15 years, people started to recognize, "Hey, it starts working." And then again, all these claims we came up with, we will have robot butlers at home and automatically driving cars and all that. And people really jumped on it. But I think it was the first time that some of the promises get fulfilled. To what degree, is another topic, but that you could, as a non-technical academic person involved, see that is something happening. So you have Siri on your phone, or if you have this incredible Google voice assistant that is completely AI driven. So when the people first recognized it and then said, "Hey, if I can use something that's smart for my business case, I will make a gazillion dollars." And everyone jumped on this early AI thing. And now companies think, "Hey, I must do something with AI."}

    \item{Five years ago, if you had an AI startup, you'd just get stupid money without any proof.}

    \item{Because hype is nothing logical. So you can see it if you really look at startup financing, lots of the larger VCs moved away from funding AI companies, or solidly AI companies. Now the hype is biotech, of course, everyone wants to be the next Pfizer. And I think that's how humans work.}

    \item{I guess the one dirty little truth is that I care a little about the fact that it's AI per se, but it gets a lot of resonance and interest when I use the word AI as opposed to machine learning, as opposed to algorithm.}

    \item{Well, at a certain point, it certainly resonated with investors. Maybe we're at a certain point in society now where it's almost overused, and so there's a certain backlash against just the general use of the word AI, but certainly for a period there, no matter what business you're in, in society, you had to drop the word AI in order to seem relevant and to seem like you're doing something important. Ultimately, again, from a fundraising perspective, from a customer perspective, so there's different stages of adoption of different technologies that generalize across a lot of different sectors of the economy. AI or whatever it is, is relatively early in [my company's domain], and so now, we're at kind of like what's called like the early adopters' phase, but I think there's an even more extreme form of that which is some [businesses that could be clients] want to adopt. It seems like they have an incentive to adopt AI for the sake of the fact that it's AI, and that they can then now say that they're using AI}

    \item{It's a check box. INTERVIEWER: What does that mean that it's a checkbox? PARTICIPANT: You have it or you don't. It's binary. They really don't give a rip of how good it is. They just want to say that you have it because it's a marketing buzz.}

    \item{I mean, [investors believe you have AI] because you say it and you can talk about the algorithm. I think the vast majority don't dig in because I think the vast majority truly don't understand it. I mean, there are some investors that... if you want to maybe get to talk to them, there are some that focus just on AI, and they have experts who know it inside out because they came from that space. And they're going to be savvy enough to know the difference. There are a lot of investors who don't. And so they're trying to catch up with the next big thing in tech and they're just following whatever the buzz is.}

    \item{I want to say I have machine learning and AI because it makes me sound like I'm on the cutting edge.}

    \item{I don't think it's a specific message. [...] I mean, listen to CNBC, look at the investors in startups. And I think if you scan the vast majority of startups, anything related with tech is going to talk about their AI engine or their machine learning engine. [...] But the fact is at the end of the day, regardless of what you're doing, the end goal is that you're meeting an unmet need, and AI and machine learning is just a way that you're getting there. So saying that you're using it in your product is the way that you solve the problem. And right now, because it's the buzz du jour and everyone wants to do it to say, "Yeah, we're on the cutting edge." So that's why you'd see more of it. To be fair, there are some investors that see it as, okay, this is cutting edge. And with multiple evolutions of this, we are going to get to that point where this starts to overtake humans in terms of their function and intelligence. And that'll happen at some point in the future. But I believe that with 95\% of the uses of AI and machine learning, it's just a way to solve a particular problem and meet a need. And it's just a slightly better algorithm.}

    \item{I have my own personal definition of AI. I'll tell you. I think it's just the marketing term for the ability to do massive amounts of equations in order to make predictions. I don't view it as actual artificial intelligence. I view it as a marketing term in order to ... The ability to use very clever algorithms to do massive amounts of statistical calculations to make better predictions. That's what I view as artificial intelligence.}

    \item{I think when we talk to most of our customers, they're quite aware of AI and how this stuff works, because our target is a mostly technical audience. We end up having this sort of conversation with mostly people from tech companies who are likely to be our customers, and they've a fair understanding of AI and how it could work. [..] Yeah, we rarely mention, the AI does it. We are AI powered, that's sort of understood in most cases.}
    
    \item{We speak about AI value but we don't mention AI so much.}
 
    \item{Because most of my experience with that is in the B2B space, where everyone is somewhat technically inclined, or at least people making the decisions are familiar with the general typology of what's out there. And they understand at least and in broad details, about the benefits that AI can offer a business. But in terms of users... Just giving an example of, when I tell new people that I meet what my job is, what I do, they're a lot of times, "What's that? I have no idea what that is." It's definitely not as strong of a marketing tool directly to users.}

    \item{So we're not fundraising right now, still bootstrapping. So I don't know if it will help or perhaps even the opposite, be not helpful anymore. When we fundraise, I think with, to be honest, most customers don't recognize it.}

    \item{If you have something that you say is a predictor and your name is something .ai, you lose a lot of credibility when your predictors are not quality. So it was just this growing pains. It happened really early on when there's probably like a couple hundred users. We adapted to it pretty quickly but it's something that the real ... When I see it every day in my cloud customers, it's a real concern.}

    \item{I would say from the perspective of founders that would be [a signal] to investors and clients as well, that's definitely not at what we do simply because this is something that the founder, [Founder Name] went to school for and really has a passion about computer vision. And so he wanted a true computer vision solution moving from the get-go. But yeah, even without the AI moniker, [compared to] before we were considered [not an AI] company, before we [developed our AI features...] there's more attraction from investors if you are labeled as a AI company.}

    \item{...one of the things I've learned about what makes this distinct is that when you're fundraising, or if you're even thinking about fundraising for something even remotely related to AI/ML and the market potential is as big as it is and you're at a time right now where the liquidity is really, really high in the market, they find you. [... ] It's a very different kind of power dynamic.} 
    
    \item{there oftentimes questions [from investors...] it's just like, "Okay, how does your AI work?" So then we have to describe the backend processes. We have questions about where does the data come from? So that's an often question. I think those are the two most common questions.} 

    \item{INTERVIEWER: So you said that you hate that AI as in your name. Then why is it in your name? PARTICIPANT: Pretty much because we want to ride the hype cycle too. So let's be honest here, of course. And I don't see that negative as I might've sounded. It helps, labels help people.}

    \item{[...] there are lots of already set up AI frameworks that you can just connect to and apply to your own products and then you have an AI product. Or other cases that I see personally most often testing different tools as a marketer is that lots of tools are automating something, and then they say, "Oh, I have an AI."}

    \item{I actually wouldn't classify any of it [as AI]. Anything I see today I think artificial intelligence is not applied because none of the algorithms are self-aware. So I actually think that the whole... You see this a lot in tech. I'll use autonomous driving as an example, right now where it's like, "Oh, I got full autonomous driving." Bullshit. That's two decades away. I mean, anyone who really can look at those algorithms and see what's happening without having some sort of sensors in the road, we're so far away from that and all these edge cases to get there. I mean, same thing with AI. People hear of AI and they think of robots or the character in the Marvel movies who's actually a fully humanoid, thinking being. No. And I could be wrong on this because it's not my area of expertise, but I just feel that there's so much... Right now it's one of those buzzwords that everybody's jumping on to say they have it, but very few actually do. In reality, what it is are better algorithms to figure things out. And to be fair, I mean, there are some things that are really good, like the voice recognition. And if you think about the AI like with Siri and Amazon's Alexa and the amount of language processing that's happening to pull things out, I mean, that's very impressive. But at the end of the day, they're not self-aware. Yet, I hope. I mean, I don't know. But at the end of the day, it's just a really good algorithm. And so I think a lot of this people are glomming on to that futuristic view of it. And it's that next big, futuristic thing that they can do. And I think it ranges from, oh yeah, we're using machine learning and AI. On the one hand it could be just it's an adaptive algorithm for something fairly simple that's looking at a relationship with two variables all the way up to now we actually have a whole platform like Siri and it's a different thing. And that entire continuum contains AI. So everyone wants to say, "Yep, I got AI. I'm just like Apple or I'm just like Amazon."}

    \item{I'm so bad with names, but if you look into the acquisitions of Salesforce from the last three years, and I think it was 12 companies, 11 claim to do AI. And Salesforce bought them and with big marketing, "Hey, we bought another AI company to do our Einstein platform," I think it's called, "To make it better, smarter, faster." And after half a year, if you looked into it, they just discontinued all the companies because you find a little press release or if you know someone who works at Salesforce in San Francisco, and once again, they couldn't do what they claimed it could do. So I know that it's very episodical. INTERVIEWER: Yeah, but it sounds like then from your experience, a lot of the companies that are claiming to do AI are telling their customers, they can do AI, ultimately those solutions are failing. Is that accurate? PARTICIPANT: I don't know if they are failing, but at least they're not succeeding with AI technology I'd say.}

    \item{[...] Artificial general intelligence. They think of something that actually can think like a human being can think, but an AI model doesn't actually think right? It's just, it can maybe make you, fool you into thinking one day, but it's actually not really intelligence. It could do massive amounts of calculations and use that to make predictions. It's not the same thing as actually, intelligence. You, as a human being, I could suddenly tell you something completely unrelated and you could apply what you learned to figure that out. Is there an AI model on the planet that can really do that? No. Could you walk and run, and suddenly I teach you about, tell you a little bit about what a movie is, and watch a movie, and you'd make comments on a movie? That's real intelligence. A human being has intelligence. What's amazing about the human brain isn't so much that it can do things, one thing at a time. It's the fact that it can take things it's learned in those things and apply them to completely different situations and come up with its own ideas. It's not like an AI suddenly is going to come up with a brand new statistical [domain of application] model for you. That requires, still, human intelligence. That's what I view as true intelligence. That idea that creativity, that an idea suddenly pops in your head and then you can implement it and come up with something new that does not exist. [...] To me, that's real intelligence.}
    
    \item{I tend never to use the word artificial intelligence. Internally I think we almost exclusively use the word machine learning. I think the FDA now kind of labels some of it as AI so I use it a little bit more now publicly but there was a time where I probably only used the word machine learning because AI feels so amorphous.}

    \item{I think absolutely, when anybody hears the term artificial intelligence, they're thinking of HAL from 2001. They're thinking of all the science fiction novels they've ever read, or all the movies they've ever watched, or the Terminator from Terminator 2. I think they're thinking of something that mimics human behavior, that has a consciousness. INTERVIEWER: Some people would think that evoking something that, or using a term that evokes The Terminator would be a bad thing, would not be a good marketing tactic. PARTICIPANT: I think that's what they think. It's like it can actually replace human intelligence. But they can't. I do not see any AI model that's even close to human intelligence right now. Even a child's intelligence. Even a child's intelligence, I do not see it.}

    \item{So we've been around [for several years]. I think there was some pitch competitions in the beginning when we were a small startup and I remember there was a few other companies that were pitching and a few of them were using the term AI and I remember just listening and realizing I don't think there's any AI at all happening. I can't quite remember the application but I remember being kind of annoyed that they got to use the hype term whereas I didn't really think there was any of that happening.  I think that's probably less so now. I think a lot of companies really are using machine learning more than they were five years ago because if you have a lot of data, that's the right thing to do.} 

    \item{So I definitely think that in the same vein as like crypto and blockchain, as buzzwords, as companies will spin up the idea of providing a service that can be automated with artificial intelligence and what they wind up doing is doing a bunch of manual work to make it seem like they can provide that service. But once they go to scale, it doesn't scale very well because you're still doing a lot of things manually and not doing a AI data driven approach first, because from day one, we started with a, our proprietary model that we began training to make sure that we weren't like, "All right, well, submit us [the raw data] and then we'll process it and then give it back to you." That kind of thing, because it's just not a scalable solution.}

    \item{But in terms of why we decided to include a bunch of real AI, deep learning, all that kind of stuff, it's the only solution that will enable our grander vision}.

    \item{Personally, it's a pet peeve of mine, that some companies will call their solution AI. And it's like, "Okay. You used a random forest. Congratulations." But no. There's that. You can get value. The value in the technology that you use doesn't necessarily even have to come from the technology. Just saying you use AI. In the same way that if you're interviewing, saying that there's like, "Oh yeah. I know like ML, I know these ML frameworks, whatever." Even if you don't really. It's a huge bonus. I think the value we deliver is like, "We do use AI/ML technologies. Here is exactly where we use them. And then here's how it's making your life easier." And being able to actually substantiate that with results from our application, I think it's more than just a marketing edge and it substantiates the claim.}
    
    \item{On [a recent date], I did a pitch event. It was a virtual pitch event with about five or six investors and an audience as well. It was a three-minute pitch. I ran through a slide deck, talked about the core technologies and the products that we're building. After that, there's was a two-minute QA where the investors asked me different questions. So from that, four out of five of the investors reached out to me to say, hey, I'm interested in what you're doing. So I said, hey, thanks for your interest. I'm not currently raising funds, but I'm more than happy to keep you informed about the progress we're doing. Here's some materials you can read over. We'll reach out to you every month with our progress.}

    \item{The reason we use the word is just it sounds cool and people like that in the marketplace. Like, "Oh, you're going to do AI on my data." I'm like, "Well we're going to do machine learning on your data." I don't really have the need to do AI where I would with these kind of things, right? It's not that complex. I'm sure someone could. Like if you're trying to say, "Hey, I'm looking at this data and maybe can I impute whether you have an illness or something?" That probably would take something a little bit more nuanced and AI-centric. But that's not what we're trying to do.}
    
    \item{So our overall strategy is to take as little investment as possible. And there's two sides to that. One is to keep the burn as low as possible, two is to get to revenue as quickly as possible. And we began with seed investors. We actually did [a University angels funding event], that was some of our first money in. We had great experience with a crowdfunding platform. Easiest fundraising I've ever done in my life. We raised a [an approximate sum] in \$10,000 checks in one day. One day, I couldn't believe it. And it made me look at these things differently. It allows you to control your destiny a lot more.}
    
    \item {There's angel investors. There's VCs. It depends. I can't pick the kinds of investors that come to these [pitch] events. I only know the panel maybe after I sign up, but it's a good mix. It's fairly classified angel investors, accredited angel investors. Only those accredited angel investors can actually invest in startups. There's a key distinction. Not everybody can give you money. I have to say no to a lot of people, actually, because we don't have enough of a net worth to do business. But if they're accredited investors or venture capitalists, we'll keep them on our Rolodex.} 
    
    \item{I always try to speak to the dumbest person in the room. Based on the questions, I'll go into the level of detail that's appropriate. I never like to start up here because you just lose the audience immediately. The biggest barrier to data science is being able to tell the story. I think a lot of data scientists really are not good at that. So if you have really technical founders, they don't know how to relate their product to laymen, and so I always try to be cognizant of the fact that I'm talking to people that are not industry experts. They're experts at determining value propositions, and so I need to be able to accurately say what my value proposition is.}
    
    \item{So we're a venture backed company, so we've raised a Series A in the past, and as a company we have to continue raising. So our funders are typically other venture capitalists, we talk to some strategics, so other big companies in the area that are really interested in what we're doing, but kind of want to play an observer role. So we talk to them and then we also try to make use of government funding as well, so there's some kind of programs available for non-dilutive funding for grants. So we make use of those three avenues of funding.}
   
   \item{I think we certainly want to make sure any VCs that we work with are well-aligned with us and can provide added value. We're not just looking for source of funding but also some support and oftentimes VCs or funders, equity funders will take a board seat, and so we want to make sure we work well with them. So there certainly is a mutual vetting process any time that there's a potential relationship in the works. But in general, the VCs that we work with tend to be in the healthcare space and be from backgrounds that could help us, they have a lot of contacts or they know how to build these types of devices or how I think through pricing models, commercial strategy, things like that.}
   
    \item{So we're funded through investors. So [Company Name] has done a seed round, raised about [an approximate sum of money]. And so we're very lean. I have done venture backed businesses. And I actually try and avoid that business model because I think oftentimes, the VC's goals are at cross purposes with building a long term successful business.}
   
   \item{And [at a previous startup] the VCs wanted to hype things up, get a lot of press, make a splash, so they could raise the next round at a higher valuation and look good to their LPs, which was actually contrary to what we needed to do for the slow growth to build the business. In [current startup name], our goal is to get to revenue and cashflow positive as quickly as possible. In many ways, VCs, they won't say this directly, but as an entrepreneur, once you have revenue, it's problematic from a fundraising standpoint. Because before you have revenue, you're all about promise and potential. Once you have revenue, there are metrics. And the question becomes, why aren't you growing faster? And it's very rare that something comes along with the true hockey stick growth that VCs are looking for. So that actually puts you in this mode where you want to go on the hype, as long as you can, put off revenue as far as you can, which puts the entrepreneur in a defensive position, because the only option then is to raise VC money for the next round, and of course, the VCs want you to spend more and more, because that's what their metrics to their LPs look like. So I found that for many, many businesses, it's not the right model. There certainly are cases where if you talk to Peter Wendell at Sierra, he'll say, "I sell rocket fuel." If you're not going to Mars, you don't need rocket fuel. There are actually very few businesses where rocket fuel is the right thing. So I'm mindful of that, and that's something I've learned. INTERVIEWER: It sounds like that can create a vicious cycle for the entrepreneur in terms of interrupting good product development. PARTICIPANT: Totally. And the burn gets higher and higher, your runway gets shorter and shorter, the expectations diverge from reality faster and faster. Yet the entrepreneur is in a situation where, I'd feel this very viscerally where I'd been one thing to my customers, they care about what I'm doing today. And my investors would care about where's this going to be in five years? And as that gap got bigger and bigger, that's a huge source of stress.}

   \item{VCs are doing just fine these days. And it's also like, where people are in their careers and their experience levels, there will always be a pipeline of people coming fresh out of college early in their career, where they want that rocket fuel. And they don't realize that it's only going to work out for 5 percent of them. And as you get later in your career, you realize there's a lot more options than VCs. There's other ways to build a business. Whereas when you start out, that seems like the, maybe it's the way we're taught, but that seems to me like the only option. I didn't even know there were other options besides that, to build a business.}
   
   \item{I personally do not want to pursue funding just because then it will be like a real job. Like I'll be in debt to someone and someone else will influence what we're doing. If I didn't work every day, that might be interesting, but it's not interesting right now. It's like an endless stream of people who want me to go present and talk to people but it's like if I do that, what's the benefit to me? It could be useful to solve some problems, but I think it adds a lot of complexity to where it's like all of a sudden, "Eh," like I'm having to spend several hours a day on this as opposed to if I'm working hard at [my main job], this stuff goes on pause. If [my main job] stuff is chill, I can work nights and weekends on this. That's the other thing is the more interesting more for me is to have an equity partner who is really good at user experience and maybe one who is really good at data science as opposed to someone who just has deep pockets, right? Then it's all we're doing this for our passions and what our shared interests are as opposed to trying to get rich. Maybe some day it will be something I could sell to someone if it has a large user base and has a proven track record of profitability, that's probably interesting. But until then I don't need somebody to give me a bunch of money.}

    \item{The best investors and the most helpful ones have really been staying up to date with what we're doing, and we have a number of asks that we ask of our investors and we have great investors who will follow through with that. And it really depends on who they are, so if we're looking for advisors in a specific, like [one aspect of the industry domain], we know who to ask. If we're looking for advisors in brand building or marketing, there's people to ask those questions. Or how do we think about press, something like that. And so, it's really folks that are engaged and willing to help in the areas where they're able to.}
   
   \item{I think there's not really a formula for it, it's kind of a feel of would I want this person on the board and likewise on their end, it's do I have enough belief in this company that I think they're really going to succeed and I want to work with these founders or work with the management team.}
   
   \item{We did it pretty simple, we put up a website, said what we have as a product and looked who signed up and then went from there. And yeah, we had both pretty well-paying jobs before, so we didn't have to charge a lot, we wanted just to get iteration speed and did that. And then we somehow noticed, "Hey, we now have X customers and the company is running itself more or less." And yeah, so we stumbled into it.}
   
   \item{So, in our case, we raised pre-seed. So not yet to seed. I think that's still something we're working on. So for the pre-seed, we applied for an accelerator program. We actually applied for a few. And then luckily we had a choice at the end, so that was great. But yeah, we basically just applied for a few accelerator programs, because we felt like we wanted to go through a program to get mentors on top of just the funding for the early stage, and this is what they did. That's kind of was our focus.}
   
   \item{And because I think Silicon Valley has this idea of, you've got to grow like a rocket ship speed and huge margins.}
   
   \item{So we had some great conversations; a few investors who actually were interested in joining, but I didn't feel like they were in line with our own vision, so we didn't agree to proceed there. INTERVIEWER: What about them made you feel like they weren't aligning with your vision? PARTICIPANT: Yeah, so like one of the very prominent examples is that a VC's focus specifically to just do the seed round and then make sure that the startup doesn't get additional investment and just exits, so basically sells the company right after, which is not what we want to do. I want to develop our technology to have a [the fully developed product] for everyone, and then go for an IPO.}
   
    \item{But I also don't think we would have built ML if we didn't have VC money. And I think you're right. We would have probably gone to more stable, "Let's build a good [core services business] with good software. And once that is on sure footing, then let's maybe some special projects."}
   
   \item{This is something we actually see with our own customer base. We recently had a fundraising round, and instead of going through VCs, we've generated ... It's everybody who's invested in us uses our product. It's very simple. You come in here, you see the recommendations we get. You try them out and they work, and you're like, "Oh, wow. This is legit stuff. This is very, very legit stuff." You know?}

    \item{What maximizes efficiency, and what's something that people are willing to pay for? And oftentimes it's efficiency. I'm so cynical. I don't think it's quality.}

    \item{So the biggest thing about ML is like, oh, take a step back. Software is pretty cheap to make and run. Other than your engineers, the most expensive part. But for machine learning, you have to label a lot of data. You have to label thousands of images for your training, for your validation. And then you have to pay [specialists to label your data]. That's what's hard. You have to pay people who are [specialists] otherwise then you get bad ML.}
    
    \item{So angels, we consider any individual that's not part of an institutional fund. And the strategic part of that, are really people who have subject matter expertise in something having to do with our business. [...]. And so, we have angels that really have that diversity of expertise in those areas.}

    \item{INTERVIEWER: What do you think motivates the questions they ask you about your modeling approach or data approaches? PARTICIPANT: Revenue generation opportunities. INTERVIEWER: Can you tell me more about that? PARTICIPANT: I would say first and foremost, investors are investing in a company because they believe that it could grow and scale, and ultimately exit and provide liquidity. So, the questions or conversations that we're having from a business operations, whatever standpoint, is all thinking about how are we building the best business that we can, so that we can grow as quickly as we can to create some sort of liquidity for investors? And so, I would say that's the basis of the questions.}
    
    \item{I would say both because institutional investors that want to invest in AI companies that are very mindful of how they're applying AI and doing the thinking for the investors essentially being like, look we are aware of all of the problems at large and upcoming regulation and all of the confusion around it. And we want to stay ahead of it and educate people about it and do it in as transparently as possible manner to make sure that people are comfortable with the solution before it is deployed en mass.}
    
    \item{We're VC backed, which means that we believe that we can IPO at some point. If the IPO doesn't happen, then there's a number of strategic exit opportunities that would make sense for this kind of company.}
    
    \item{We've had private investment and then we also have a crowd funding round that we raised. I think we're almost at [a sum less than 1M], anyway it's on [a publicly available site...]. But that has a lot of our publicly available information as to share count a number of investors and all that kind of stuff.}
    
    \item{So we went the route of equity crowd funding, so it's a little bit different than like Kickstarter or something like that. Yeah. And so one of the main benefits is we're like, "Hey, if you invest in the company, then you own a share account. And so there's potential for you to have a return on that investment." And that's the major selling point of doing equity crowd funding like this. And was it really only possible because of newer federal laws that allow that kind of funding. The main benefit was just a little bit more transparency to individual people. And we learned a lot from it. There's a lot of people that we talk to that are business sales, engineering, whatever. And they're like, "Okay, we'll take this to our leadership and talk about doing business with you, but at a personal level we would like to invest."}
    
    \item{Yeah, I would say the downside of it is just that funding, it trickles in and it's less lump sum payments, but it really just depends. The main downside that we face through this equity crowdfund funding is through the company that hosted it and not necessarily the fact that we did crowd funding because we are interested in doing crowd funding again and giving that option to individuals, to invest in the company without being an accredited investor. INTERVIEWER: Why do you think that's important that part about not having to be an accredited investor? PARTICIPANT: Well, because there's a lot of people who do not meet that classification and we are very much in a market where people are more educated about investing options. And even though you may not have the assets or capital to do what an accredit investor can, you can still take a portion of your income and make small investments in a lot of companies. And it gives people more options to invest in companies that are not publicly listed yet. So I think it's more of a play towards what the wider audience is interested in and then also in the spirit of transparency and just saying, "Hey, if there's people out there that don't like this, we have a public page that you can express your concerns on that we will address that are on the same page where people make investments." So that way you can get a good picture of where we stand on happening events or things that we've addressed in the past.}
    
    \item{[The reason for the chosen funding path] is that because we want to continue bringing on advisors that will help guide us and that's what good VCs do. There's lots of good VCs out there that we could partner with but we're an unproven entity and just moving forward with [a specific sum of money] isn't going to get us there quickly and that means that my co-signer and I would be constantly having to raise another [a specific sum of money], another [a specific sum of money], and constantly fundraising rather than getting out there and running the business and building it and helping people and that's one reason why the philanthropists have started [this funding entity] because they noticed that women and people of color in particular, they want to help their communities and that's one of the reasons why they don't grow as quickly and scale as quickly and they don't get the funding that they need in those series of rounds because they're so focused on their business and giving back to the community rather than just growing and scaling so they can get to IPO.}

    \item{So it's self funded. We're on our own, we haven't borrowed any money from anybody. Nor are we funded in any form. So we are on our own, entirely.[...] So the fundamental belief has been to expand organically, to grow organically. So that's the power of the organization. Extremely hard working bunch of people who believe in each other's abilities. We believe the CEO's vision. I don't think that we have needed all these years. I think our biggest goal is to become self sufficient. To make sure that we recover the costs that we invested in the product over the last several years. We try to recover that over the next year or so and make sure we grow much bigger so that we redeploy the funds back in the product and come up with more [AI] solutions.}
    
    \item{What our CEO is anticipating [from a crowdfunding round] is it's likely going to come from [Company Name] super users; people that are really happy with our platform, see the direction that we're going, and wanting to be a part of that.  Honestly, we feel like the features that we're building out are going to be so useful to people, that they'll have no problem investing via a crowdsourcing way, at least some percentage of our users. Just because it'd be like, "Yeah. This thing is great. This is free money. I invest now and I get... Because these guys are obviously going places." That's what we hope our users feel, and that's what we think we can offer.}
    
    \item{We have investments from private individuals, and while we were trying to raise funding, there were some investors that they really cared about the fact or that we are trying to take an ethical approach in what we are doing. They really, really like that and supported that. There were some other investors that they were turned off by it. You see both. I heard by at least one person saying it out loud and more than one implying it, saying that give it for free and just get all the data and sell all the data, which is not what we want to do, but this exists. For the most part, we are trying to raise capital from investors that have the same kind of values and mindset with us and people who are not afraid to lose certain revenue or sales ... We had clients asking us to do things that we said, you know what? No. No, this is not something we feel comfortable with doing.}
    
    \item{I don't know if you're aware, we had done an equity crowd funding round that we were allowing anybody to go and buy shares online. Part of the reason we did that was because we wanted everybody to know that you have equity, the crowd, all people, we have [specific number] investors, an average investment size of [specific dollar value, less than \$1,000]. Everybody could go and invest. We didn't just do it behind closed doors to give the opportunity to very few people.}
    
    \item{So we formed a huge network of venture capitalists and scientists who are now convinced or investors and so forth. And a lot of CEOs who ended up mentoring us during the process, we are still in touch with. And so that was a really good experience. And although the feedback used to be really hostile, oh, this is not going to happen. I mean, the early days, I think we had a prototype and some people felt it looked like a high school project, until you could really convince them that we had data and we were doing all these things. And so, it sort of took us and people can judge you really quickly, because they get like a five minute pitch to judge you. So it also tells you, one thing we also learned was that you might get judged very quickly, but the onus is on you to prove otherwise, that you have to really battle hard and show them that you are doing that. And maybe it exposes certain things that you're taking for granted in terms of how you're pitching your company.}
    
    \item{So, we are bootstrapping and we are going to go with a debt arrangement and the reason we're going to go with debt arrangement is because we know, based upon our research, that in order to capture market, in order to get the resources we need, in order to maintain a competitive advantage, we're going to need to grow and scale quickly and that's going to require more capital than angel investors could put up and we're going to need to do that faster and have more control over that process than if we go through the traditional white male VC process.}
    
    \item{We're actually trying to do a lot of our seed round crowdfunded, which is a bit different obviously than how it usually goes. I would anticipate that, do we take money from VCs? That's something that I would need to address and prepare presentations for. But for as long as possible, we want to avoid taking money from VCs. We don't want large individual stakeholders in our business. Because oftentimes they're very focused on profit, whereas we're just trying to build the best platform for people. [...] Our CEO, he's run a few businesses/startups in the past, and not having the best experiences with VCs to my knowledge. For instance, a lot of the times VCs will want a board seat after giving a certain amount of money, or they will set profit goals for you. And we've all agreed that, those things can inhibit progress toward our collective goal. Everyone that works at [Company Name], like I said, the pay isn't why people work here. Everyone that works here is really passionate about the product, we think we can build something really cool and useful to people. And that part of the vision is taken away, or dampened, if I get asked for hard quarterly profit goals, or someone from the VC saying like, "I think this is the direction you should take the app."}
    
    \item{I would say, we looked at the VC funding route in the beginning because that's what you're told to do, right? That's how you get funded. This is the path. You go pre-seed it's angel investors, after that it's VCs and then you go through the Series process. Well, if we want to not dilute our company, which our employees have shares and will continue to have shares and so if we wanted to do the right thing for them and not dilute their capital ownership, if we want to be able to have control and run our company rather than having to go through constant funding rounds, then this was one of the best alternatives we've seen in comparison to using like a small business loan since we're pre-revenue and since we would need more capital than what one of those small business loans could offer since we're a tech company and we're going to be national or an app, right? So you can download the app at any point in time today. So we need to get that go-to-market strategy out the door quickly. We need to have our brand presence out and ready to rock and roll, go-to-market strategy messaging consistent across the board and we're building a marketplace. That's really double the marketing costs.}

   \item{Yeah, we had a number of pitches that were made to tons of investors, and it all depends on who the investor is. Some investors are very curious about every slide that you put together, and some investors, they want to just know, "Well, tell me how you're going to generate the traction? What your revenue is going to be between now and three years? And how you want to return my funds back with what profit?" So there are very shrewd investors, there are very curious investors, there are very deep minded investors. So you run into all kinds of investors in the process, but ultimately, everybody is looking for, "when is your first pay day? So, let's say I invest in your firm today, when are you going to pay me back? And with how much return?" So that would be the bottom line with most investors, but some in the process will be very slow in asking that question. But some will be very fast. They'll probably ask you in the second minute, "well, I've seen tons of these and I've gone through tons of these, tell me the bottom line, explain the bottom line to me."}
   
   \item{Because maybe a lot of companies that you speak to could be B2C or B2B could be ready products, where they will have different challenges. Our challenge is not sales, or customer interest, but the time taken to implement it, which in turn impacts our revenue and stuff.}
   
    \item{Most recently [the questions investors ask are] really around, "how do we think about making sure the models are valid? How do we think about improving the models, building new models as our platform grows? So, what is the roadmap for all of it?"}
   
   \item{What I'm speaking about, these are my impressions. By vision alignment, I mean somebody who understands the larger problem that we are trying to solve and the impact that we will be able create in the coming years, and the value that we provide as a platform. And strategic alignment also means, investors who better understand this business or have been in this space, have invested in the ecosystem, not necessarily AI companies, but companies that could be our consumers, like in [specific, relevant industries.}

    
    \item{[Investors were] like, "Well, these are just science projects." But the reason it rubbed me wrong was that, that was the fundamental key to why we thought that our product would ultimately succeed on the market, because of the techniques and rigor that we used. [...] it was almost like an anti-science trivialization of what scientists do.}

    \item{The difficulty we have is that [our underlying mathematical field] in itself is really not looked at ... People look at NLP. People look at computer vision. Those are the kinds of sectors of AI that have gotten a lot of attention and a lot of brainpower, and so a lot of people have created many different packages to make those applications really quick. Now you really don't even have to build it. A lot of platforms are just click to play. You can load a data set, and you can say, "Give me some image recognition," and you've got it. With [our underlying mathematical field] and the way the technology is right now, the pipelines don't exist. The algorithms don't exist. They're just in papers, and so that's the toughest part is just going through the equations and ensuring that your pipeline is right so that your output processes through the equations correctly.}
    
    \item{So we're trying to solve a specific problem to publish a whitepaper. So what we want to do to prove the efficacy of our model is to apply our data set that already has [ground truth labels], to impute it. We've built our ML models, and we have [ground truth labels], but we want to compare it to a published industry data set to see how much better, quantifiably, our algorithms will be over what they're doing currently.}
    
    \item{The only value that they understand is the delta between my [predictions] and their [predictions], and so if I can show that my [prediction] is more accurate, they don't care how the sausage is made. [...] But just like the iPhone, you just want to know that it works, not necessarily all of the technology that goes into making it. So I always emphasize the accuracy of the [the predictions] based on what they're currently doing [...] but we don't go into details.}
    
    \item{The choice of the relevant literature is a human intervention. So, my co-founder along with our team of scientists and clinical advisors, work together to determine the relevancy of the research related to what we're doing. And then only then does it get modeled against the data.}

    \item{I'm a co-founder for this company, still very actively involved in the company and helping that development of the algorithms. But I'm also an assistant professor now here in [the academic institution] and I have my own lab, where I'm looking at these questions on the side as well. So, I'm hoping that one day these insights will further help refine my algorithms from the [scientific domain] side, more so than from the [algorithm] side.}
    
    \item{So, we have an AI and data science advisor on our board. We have one data scientist on our team, that was our first hire because we knew how integral this was going to be to ensure the accuracy of our algorithm going forward and to also leverage the data that we're collecting for academic research going forward and for other product lines.}

    \item{I'm finding that articulating these kind of nuanced data sciencey things is very challenging to the normal user. Somebody who does it for a living probably gets it very quickly. People who are end users are like, "I don't understand this." So it's one of our bigger challenges to try to make the user experience something that is palatable and easily understood. In the context of the first feature I mentioned called [feature name], a lot of people just don't understand. It just doesn't make sense to them. It makes sense to us but it doesn't make sense to them so it's not very useful. [...] You can say fairly easily the way we do it now is kind of like ... Think of it as like Z scores. Like you're either negative or positive to the norm and how negative or how positive are you and we're also thinking of like a "what if" tool to where you could say, "If I index this up, what happens to the target variable?" That I think will end up being a much more palatable experience for people than just seeing a Z-score and they go, "Thank you, I'm +1 or I'm +2."} 
    
    \item{At some point, once we get the funding, our goal would be to recruit a Ph.D. level, I'm not a PhD, I have a computer science background from one of the top institutes but I'm not anywhere close to a Ph.D., we probably need to develop the data science.}

    \item{The thing I can think of is when we were more actively fundraising and I had a pitch deck together... I came from academia, so I had an initial pitch deck that had this almost like a spreadsheet and it had a bunch of numbers on it and it carefully calculated out the return on investment and how it related to our algorithm performance and things like that. And then I presented to a bunch of my friends who were also academics and they said, "Well, this slide needs more numbers." And so, then I added more numbers and then I got more feedback from fellow like startup CEOs, and they were like, "You just need to get rid of this slide altogether." And so, there were a couple of iterations there. And so, then when I went out and actually pitched it, I pitched it to just tech investors and the no numbers thing was working just fine.}
    
    \item{Essentially the feedback that we've gotten from clinicians has been that the product that we're building really serves an area of high clinical need. So I think they're very interested. I think they're less curious about the AI aspect, that really doesn't mean much to them. Even the [domain relevant to medicine] aspect I think is not that interesting or exciting to them. They're really just kind of interested in "can you get this answer to me faster than the typical diagnostic so I can treat my patients in a more precise way?" So that's ultimately what they care about.  I've kind of found that on the clinician side there's very little understanding, maybe even very little appetite for understanding or really delving into scientific [domain relevant to medicine] explanations or really understanding the AI aspect. In fact, I don't think really anybody's asked any questions on how does the machine learning really work or which model did you use. The questions are really on the clinician impact and that's really ultimately what the clinician... makes sense that would be what the clinician cares about.}

    \item{So that's in terms of the machine learning that the team does. Upstream of the machine learning, I guess I should have said, there's the dataset and actually the domain experts, the [scientific specialists] are in charge of just curating that data to begin with. So while there's some feature extrapolation that the data scientists is responsible for, just knowing what data to get and how to represent it and making sure that the data makes sense, the [biological] data is high quality, the annotated metadata is high quality, that's something that the [scientific specialists] do since that's kind of in their area.} 
    
    \item{Yeah, I mean our advisory board has been excellent and I think while none of them are in the industry themselves, or I should say some of them are, but even for the ones who are not in industry and are academic, I think they have a good sense of... Certainly on the scientific technical side, just thinking about okay, you can simplify this or you could try this other avenue. They're very aware of the literature, so if there's something new that's come up in the last few months that we're not aware of that could really help solve a particular problem they have, they usually can point us in the right direction and they can also give us a gut check if the things that we're doing make sense. So yeah, they've been very helpful in that regard.} 
    
    \item{Sometimes we feel like the most exciting part of our product is in the details and we always want to talk about the details and also being from scientific backgrounds, a lot of folks in the company, this is their first job after academia, and so a lot of people have to wean themselves off the desire to just go straight way into the technical details.  So even though we think there's a lot of... really some of the magic is in the technical details, that's not really what marketing wants to hear, and having something that's a little bit more simple and easier to grasp is really the thing that can have an impact.}
    
    \item{Because in our field, I think where a lot of companies have failed is as soon as they go try to do that, their algorithms flop. And we had high enough confidence in the diversity of our data set that if we were to just hand it over to someone that it would just prove itself. And we didn't see that quite in the field as much, so we thought it would be a good competitive advantage.}
    
    \item{INTERVIEWER: So you said that they were disinterested in the technical details. Were there instances, specific times where it showed to you that they were disinterested in those details? PARTICIPANT: They said it right away.}
    
    \item{I think because ultimately the folks that we work with are trained and at the end of the day they're trying to care for patients and there's, I'm sure, a lot of noise that comes along with all sorts of external diagnostics and services and are really laser focused on how do I make sure that I'm giving the best care to patients. So I think there's a bit of that.} 
    
    \item{The other challenge is that ultimately, our results aren't going to be as stellar as a lot of others because now we have to account for the fact that it's for all the variability within the data set whereas, if we're just focused on one homogeneous data set, our accuracy stats will be higher. So, that has been one sticky, like difficult point in terms of head-to-head comparisons, for example, between us and our competitors. But what it does buy us is that when we do deploy to a new site where we have less of a drop-off [...]}
    
    \item{So a very educated potential customer would say something like, "Where's your data from?" And most doctors will be like, "Okay. AI, what's the error rate? How do you know if it's wrong? Who's liable for...?" But it's more about how it impacts their practice rather than explaining how the model works. [...] And then sometimes they're curious, so a particularly curious techie [doctor], we have materials I can just copy and paste into an email, and they can learn about a three-minute video on how CNNs work. And "Here's the Google paper," and different materials if they want to consume that. Because some do it, it's a rare bird.}

    \item{[...] and the second, I think is a fundamental understanding of statistics and how those statistics apply to large data sets and the different pitfalls that you can run into when you run these statistics on big data sets and how the data can in a sense, lie to you, or you can lie to yourself. It's very easy to kind of get in trouble with that.}
    
    \item{A lot of it in our field has to do with overfitting and out-of-sample validation. And so, I think a lot of people, whether it's on the research side or in the academic side or the industry side, go about building machine learning algorithms that demonstrate very high accuracy, and then they're able to publish it and make pretty big news with it. And it's not their fault either that ultimately, these models don't end up generalizing well, because they don't have access to the data sets to even validate that, but there was a big wave of these initial studies just showing numbers that are through the roof AUCs of 0.99, and things like that, almost too good to be true, but then time and time again, when you take these algorithms and apply them in the real world, or have slight perturbations in the patient population or whatever other features it is, then it kind of falls apart. And so, that was probably one of the prime things on my mind when trying to develop our algorithms because yeah, I felt like we wanted to take that next step and make sure that once we did deploy in the real world, that it wouldn't kind of fall apart. And so, we do a lot of testing in terms of trying to do completely out-of-sample testing, but also sourcing our data from as diverse sources as possible in order to make sure that we have the highest chance of generalizing}.

    \item{It's still something we're trying to deal with because I think it's a hard conversation to have when you have that conversation with people who know statistics, and then it's an almost impossible conversation to have with people who don't know statistics well at all. So, what I mean by that is even within stats, what numbers do you report? There's a lot of wide range of different values that you could report and they have different meanings in different contexts. And so, the challenge really is trying to boil it down to what it ultimately means for the performance of a certain product. And honestly, it's still a challenge we're facing now, because there are certain anchor points that people in the field have, whether or not they have a good understanding of statistics.}
    
    \item{I've been talking about this for four or five years, so it's hard for me to recall a specific thing. But then, when we've talked to even our current investors, it's just like, "If you can get above 90, that's..." Or even doctors will say, "90 or above. That's really great. That's really amazing." But I think it's just like humans are comparative creatures, and they're not good at absolutes at all. Nothing is an absolute in how we talk about things.}
    
    \item{The 90\% bar comes from the fact that well, it is the psychological effect that [...] So at a fundamental level, the 90\% is like that your Macy's sale, when an item is 9.99, you tend to look at it and buy it more. So it is from a marketing publicity standpoint, that is just one aspect of it}. 
    
    \item{Nobody cares about the exact percentage. When we're talking to most of the customers, they say is that at least 80\% accurate. That's what most people put us there. 80\%, if it is at least 80\%, I have heard it from so many people, then they will work well enough for us. INTERVIEWER: Why 80\%? Why do you think? [...] PARTICIPANT: I have no idea. I have no idea, but that's what I have heard many times. For them, it's a good confidence level to make them say, okay, we will at the insights we are looking for. INTERVIEWER: What do you think their conceptualization of what that even means is? PARTICIPANT: [...] I don't think most people have a very clear way of defining accuracy. I think it's more psychological. Can I trust the numbers?}
    
    \item{That's the number one requirement as to doing anything less than a deep learning will not give us the type of results that we're looking for. So anything at the linear modeling level will not give us the accuracy that we're looking for, at least, about 90\%}. [This quote also included in Technological Pressures section of this appendix.]
    
    \item{So, investors are a heterogeneous bunch too. So, if it's the typical tech investor, they often aren't as familiar with biostats, and so often, we don't even need to get into this conversation. Like, they almost take you at your word that your algorithm works and they never dive into the details. And then there's like the kind of biotech kind of academia-related investor, and it's almost the opposite extreme where they get completely lost in the numbers and not potentially just enough to be dangerous sometimes in terms of trying to interpret what you're trying to say. And so the messaging, there has to be much more careful and well-thought-out and airtight because any tiny gap in the logic will result in disaster there.}
    
    \item{And obviously we've got to do that whole thing and do it legitimately. So I am somebody who is wired that -- all the products I've brought to market -- I want them to be successful for what they are. So I don't want to misrepresent the product. I want to have something that's actually really a great breakthrough.}

    \item{I was trying to make the point about the whole generalizability, but also an additional point on top of that which was the fact that we had a competitive advantage because we were from academia. We knew how to do rigorous stats and we knew how to design rigorous experiments to validate the performance of our algorithms. And we had ongoing academic partners where I took great lengths to minimize conflict of interest and things like that, such that I would... Like our company would hand over the use of our algorithms. I wouldn't be involved in any of the data analysis or the data interpretation, I would just provide support for running the algorithm. I would provide some... Like I initiated these ideas and some advice on how to design the experiments, but it was ultimately up to the discretion of the academics on how to proceed with the project and how to write it up.}
    
    \item{Now, the AI is just one small part of it. The AI is just one small part of it in order to be able to analyze some of the data, but it requires [subject matter experts] and other people to figure out, how do we actually take this extremely large dimensional dataset and don't just use something random like PCA, or multidimensional scaling, or some other ad hoc, one of these techniques to reduce it, but how do we actually reduce it down to smaller, trainable datasets based on the [a scientific domain]? How do we make statistical models based on [scientific data] that are more informative, so that when you feed it to your machine learning algorithm, you're more likely to be able to discriminate between, make better predictions, at the end of it? To me, these challenges are actually real opportunities because of the fact that we're not an AI company.}
    
    \item{And in terms of the process itself, I think [these investors] kind of respected the validations that we've done with a lot of different partners and they themselves also sought to validate our algorithm independently as well, and we did that. And so, I think in a sense, that played well into our hands in that we were able to just show the product as it was without having to compete with others with a lot of marketing wrapping around the product itself and having to compare on those terms.}
    
    \item{[The product documentation is] very, very detailed. It has references up the wazoo. Anyone who has even ... You can tell a significant amount of time was made making this product. I don't know how else to explain that. If you play around and follow all the references in there, you realize even the writing itself, this is not like you're reading a Healthline article that was put together by some person in Aflac. It's almost like a review paper that would get published in the literature, put into laypeople's language. We typically look at like hundreds of references for each things we make. This is heavy-duty stuff. [...] It is quite clear a team of scientists worked on this, right? [...] People don't read the references but the fact that we put them in there, they're like, "Oh, no. They're not making it up." Now, some people do read some of the references. There are people who are into that kind of stuff, but the fact that it's there and it's explained clearly what is going on is, people get that.} 
    
    \item{[The market doesn't] focus so much on the scientific rigor or validation, but more so on what other people are doing and what they see in terms of marketing.}

    \item{You see the flow of the science in there. It reproduces the work of a healthcare provider because we make sure that what we're putting is something a healthcare provider would tell you if they understood the [biological] data.} 
    
    \item{We are basically just trying to show that we're a legitimate scientific company by having published data. We publish our AI models. We're going to try to publish these things, get them in nice journals, have that kind of validation data. Like, "Here's our data. Here's our datasets. Here's how we used it. Here's how accurate they are."}

\end{itemize}

\subsection{Technological Pressures}
Quotes are derived from P4, P23, P3, P11, P13, P9, P14, P20, P22

\begin{itemize}

    \item{I'll make a general statement. Just given that we're trying to do things quickly and we don't have a lot of manpower. When I can, I try to leverage anything pre-trained. Just because I'm usually the only one working on ML at [Company Name]. Although I do have a one intern right now. A decent amount of stuff that we use is pre-trained, but also we're doing enough things that are specialized. For the [feature] that I mentioned earlier, I wrote that completely from scratch, and it took a decent amount of time. They really just depends on how specialized the use case. But the general answer is, we use pre-trained whenever it will afford us the accuracy that we feel we need from that model.}
    
    \item{I'm still using classifiers and I do feature selection, do training, testing out of sample, that sort of thing, but not really anything, more sophisticated than that. So, in fact now we're actively talking with another group that we know really well from seven years. So this is in another [industry lab in the scientific domain]. So, they've done a very, very good deep learning based algorithm for [physiological recording] devices. They've done really well and they've talked a lot about their work, and we are in talks for a good sort of collaboration as well in the next year or so.}
    
    \item {We use as much pre-training as possible.}
    
    \item{We are training in-house. That's the primary way of doing things because to make everything [work with our hardware requirements], it's not easy.} 
    
    \item{But I haven't really learned that much after that, since we're using some of those tools. I know probably a lot has developed, the other side there's a lot of big data, they're developing deep learning models and other like that, but [I] haven't really gone that side, [I've] stuck to classifiers and regression.}
    
    \item{I would say more so what [deep learning models are] capable of doing. If you were able to get the same results with non-deep learning techniques, then in my mind, you're not using AI and that's totally okay. Because you shouldn't be using a technology just to use it, you should use whatever best fits the problem. I guess just more so the capabilities, in general. You choose the tools for the problem, and if the problem is complex in certain ways and not complex in others, and the data is there, then deep learning is a really good choice. There are a lot of caveats and there are a lot of different methodologies. Actually here's my counterexample, there's a lot of reinforcement learning techniques that aren't based on deep networks that I would still consider to be AI. It's hard to give a blanket definition, at least without thinking of it a little bit longer.}

    \item{So that's why at this point, so I'm more content if somebody in the lab wants to learn [deep learning], or if a collaborator is really good at it. So I have some people are very versed in RL based methods or deep RL based methods and using some of the other approaches, so I just talk to them and see if there are ways in which I can collaborate with them on those applications.}
    
    \item{[...] we have our own Python-based models that are ... We're not doing any deep learning or anything nutty like that, because it's fairly rudimentary data science I would say. Stuff that's accessible to someone like me who's not like a Ph.D. I'm sure there's people who would do it differently.}

    \item{The term is ever evolving, because back in the seventies you have these just logic engines, these rules-based engines that are lots of "if" statements, and that was AI. [...] The most concise answer I can give you is just deep learning. That is almost the new cutoff for AI, in my mind at least. Not to say using random forest or gradient-boosted decision trees, or whatever you want to use, isn't data science, because it totally is. I like to say AI is very nebulous term, but in my mind deep learning. These use of deep learning, is probably it.}
    
    \item{I'd contrast [current approaches to deep learning for computer vision] with, I'd told you about some of the computer vision we were doing [in our previous startup], where it's like, okay, you're trying to generate a signal in meeting some expectations about what the data looks like, about how you're going to identify things like doing background subtraction and edge detection and comparing this frame to this frame, and where you're designing and architecting the whole thing based on your deep understanding of the problem. And those techniques, versus this magic of ... I really do view these highly evolved networks as magical, that they know these things, they're in there. And we don't actually know how they know them. People are working on that, but they're more of a black box.}

    \item{For instance, some of those pre-trained models are massive, and right now we're trying to monitor tech costs because we haven't raised our seed just yet. So we want to keep everything as reasonable as possible. And if I were to just say like, "I need real-time online predictions from BERT." That's not really the most reasonable cost metrics.}
    
    \item{That's the number one requirement as to doing anything less than a deep learning will not give us the type of results that we're looking for. So anything at the linear modeling level will not give us the accuracy that we're looking for, at least, about 90\%.} 
    
    \item{So, there's still tons of work to be done. But this is what, in a nutshell, we plan to provide. But in the process, we're also employing a number of technologies like AI, artificial intelligence, the machine learning and deep learning aspects of mining the data.}
    
    \item{INTERVIEWER: [...] does your solution, is it a custom solution or is it a combination of pre-trained existing models like GPT-3 or something like that? PARTICIPANT: This is pre-trained models, no training required. Absolutely zero training, but you can have some impact [for customers]}.
    
    \item{But in terms of why we decided to include a bunch of real AI, deep learning, all that kind of stuff, it's the only solution that will enable our grander vision.}
    
     \item{So when I did [Previous Startup Name], the computer vision was super hard. And it was all classical computer vision, background subtraction, feature identification. And then I came to this, and the state of the art had advanced quite a lot between [the previous startup date and the current startup date]. And I was blown away by how easy it was with transfer learning to make [a classifier relevant to the problem]. It literally took a hundred labeled images, to get something that was greater than 90\% accurate. And that got us off the ground. So we've deliberately shied away from hard problems, to look at how we can build upon the state of the art, all the advancements that are being driven by the Googles of the world.}

    \item{I would say, providing meaningful inference from data with really high dimensionality is a characteristic of deep learning that is pretty unique to it. What else? Unique to deep learning... I can give you examples of a lot more specific use cases. For instance, if you want to do sentiment analysis on text, if you want to [task we employ pre-trained models for], all those kinds of things, those are sufficiently complex problems that deep learning is the only reasonable answer, if you are hoping to get near state-of-the-art accuracy.}

    \item{But we are hoping in the next one to five years, we will be able to provide a solution that will be almost 100\% accuracy [at the ML task]. So that's where we're going to use the AI and deep learning methods.}

    \item{At the back end, we have a mix of deep learning, machine learning models that work together, and there's a lot of NLP and NLU [components], that kind of stuff that happens in the background.}

    \item{But overall we believe pre-trained models are good enough, and custom training is a lot of pain that we don't want our customers to go through.}

    \item{It's basically feature identification, doing transfer learning on top of established things like YOLO and Mask R-CNN and stuff like that. And then occasionally doing regressions.}

    \item{So the people I'm working with are not CV researchers, who build up networks from scratch. They're expert users who know how to do transfer learning, know how to train something, maybe with a little oversight from an academic, that's a model that has worked really well. But what I've seen in that is there is a huge difference in productivity and capability. Even among people who have similar credentials and you'd think would do the same. And it boils down to I think, making a few right choices, like choosing the right net to build off of, knowing how to clean training data to get it to a good place, and being hands on in the whole process.}

    \item{I think a convolutional neural net is central to [what AI is] or some learning network. There's a training process where you learn from data. I'd contrast that with an analytical process where you build up a model of how you believe the world works, and then fit the parameters to it. And I think of, these terms are all over the place these days. But I think of it as hooking on to the advancements we've seen, like building off the ImageNet databases, building off the work in Transformers on natural language processing in that whole area. These areas where there's like a scale component to it, the more data you get, the more CPU you can throw at it, the better your result gets, provided you run the controls the right way.}

    \item{INTERVIEWER: So [deep learning is] accomplishing many of the same goals as something that deep subject matter expertise might previously accomplish and now it can do it in an automated way? PARTICIPANT: In an automated way and in a better way. Yeah, I think for me, there is this aspect to AI of opaqueness, where people work on those. They try and understand what features is it picking up and so forth? But if anything, it's like archaeology and going back and try and understand that after the fact. Those aren't necessarily designed in.}
\end{itemize}

\subsection{Normative Pressures}

Quotes are from P1, P2, P4, P5, P6, P7, P8, P9, P10, P11, P12, P13, P15, P16, P17, P20, P21, P22   
\begin{itemize}

    \item{[...] if you had this type of thing, like for instance, I personally don't want to do any work in China because you get this type of thing in China, it would be great for surveillance. INTERVIEWER: So if you don't want it to be in China, how do you prevent that from happening? PARTICIPANT: Well, we can't. I mean, I can do it by sales and by where we develop the product itself. Eventually it'll be out there. I mean, they're working on it already. And I think it's just showing the different ways that we as a company aren't going to do that. And the other thing, too, is making sure that everything we do and all the data we collect is private and it's kept private. So for instance, our data platform is going to be HIPAA compliant. We are not going to use data for any other reasons other than just improving the health outcomes and the accuracy of the product. So there are lots of ways that we can do that.}
    
    \item{So we've noticed [the idea of AI] certainly resonates with international [clients]. It increases our response rates by saying, "Hey, we're building an equitable AI," so there's a partially marketing component, but also just there is a lot of news about, "Well, what if AI actually perpetuates healthcare disparities?" And we think you can actually do the opposite. I mean, and most people, I think most people who know AI think you can actually do the opposite. You can help doctors who would have been biased become less biased if you train your algorithms right on representative data.}
    
    \item{I think one of the serious things that we think about a lot, and this is a huge can of worms, but as with any industry, there's also this negative. AI can be viewed in a negative way too. And in medicine, it's often viewed as a threat to the livelihood of physicians and their ability to practice medicine in the future because they might get replaced by algorithms. And so, that's something that in the grand scheme weighs heavily on [us]. I don't know how much it affects our day-to-day operations, but it will. There are ways that I guess it affects our day-to-day operations in terms of how we market certain things to certain groups and things like that. But it does play a huge role in just the fundamental philosophical question of how much do we aim to do, how far do we aim to take it, and what our ultimate goal is. [...] So, it's a nuanced game because when you talk to investors, you want to talk about the upside, the optimism, the hope and the goal of where things will end up. And sometimes, that's obviously way more advanced than where we are now, but that level of advancement is potentially threatening to another group of people who aren't investors, who are potentially our customers or people who to me personally, to other physicians who work with us. It's easy to talk about what we do today. It's harder to talk about what we do in five to 10 years to different groups, without being a little bit more cautious about how we convey certain things, especially since they're very speculative in nature to begin with}.
    
    \item{All this Black Lives Matter [...] I really find it hard as an immigrant to this country after half a year of forming opinion and voicing opinion about things that going on here, because I don't know the history, I don't know what unspoken things are going on in different groups in society. And I don't want to say that we don't have problems in Europe or in [European Country] with racism or stuff like that, but it was never racism based on the color of your skin, at least in the last 80 years.}
    
    \item{So we have discussed how we can offer our platform to nonprofit organizations. And then maybe startups who don't have the funds yet, but definitely need our solution. So we actually plan to offer that via accelerator programs. So for smaller cohorts, startups could apply and get our solution for free to use if they're approved.}
   
    \item{We had privacy built in by design from the start. And I think that gives us perhaps a little head start about the competition, especially here in the US.}

    \item{So I personally think that race is nothing that determines anything important that is in our [product]. So of course, on the individual level, if you grow up in an impoverished part of town and all that, and I know that [people are] ethnically distributed oftentimes in the US, so of course, that depends on that. But if I hear what the Black Lives Matter movement is citing, what has been happening in the US, I think in the 21st century, that should not be an issue in any society if it's here or anywhere else on the world. So if that is still a thing that is happening, then of course a society should discuss it and be better than that, just be better. So don't be assholes. And on the other hand, if I read something extreme, radical left-wing, it's the same, so don't be an asshole. And that is our guidance. So do we think that there are A-holes that want to use our technology or not?}

    \item{INTERVIEWER: Does your team ever talk about the ethical or social implications of your machine learning or technology? PARTICIPANT: Not that much. Not that much. We talk about it a little bit, but not that much. Mainly, for a couple reasons. [...] One, in order to get a health product, I'm going to have to make sure it's not biased by some other AI. Because it's not the same thing as one of these AI models that predicts people that have done crime and it turns out it's looking at Black people [...] The beauty of our industry is we can't get away with that kind of stuff. [...] We definitely spend a little less time worried about that because of the fact that we have to address those things before we ever get big anyway.}
    
    \item{INTERVIEWER: Have you gotten push back from [industry experts] that are saying, "Hey, there is this guy that would have paid for me but then he got [Company Name Product] for free"? Is that something that you hear from [industry experts]? PARTICIPANT: We are waiting for that. That day will indicate the fact that we did a fabulous, fantastic job, although we don't want that to happen. I'm not trying to be cheeky. I'm not trying to be give you the wrong impression, but would really mean huge to us because we broke a hegemony. We broke a tradition and we bent rules there. Let me put this on record. [Company Name] is not out to drop [industry experts] off their jobs. [Company Name] is out to help people [perform the task of industry experts]. There are people living in [some parts of the world] who live on less had 30 dollars a month [...] there are people who spend an entire month in that budget. But then all they need is an internet connection and maybe they borrow a friend's laptop or a computer, access to it and they can [perform the task of industry experts] on [Company Name Product]. [...] So the day [an industry expert] says that "these guys have robbed our jobs," we'll definitely want to invite them and have a cup of coffee.}
    
    \item{So technology wise, I have the utmost respect and admire what Google is doing when it comes to artificial intelligence, or what Facebook is doing, especially Facebook in the last year [with their open source technology]. When it comes to ethics in AI, I think we all witnessed in the last couple of months, how Google failed. Especially for tech, I'm really a left-leaning liberal, but I oftentimes think back to university... there were some [people], a little bit older, that were just males in computer science and math. And of course, if you're strictly with other males confined for a couple of years in some cellar with computers and all that stuff, you have a distinct way of perhaps talking, making jokes and fun. And I know that these jokes are not an appropriate representation of our society. And I think that is oftentimes a problem with tech companies, these "tech bros", here in the US. I don't want to defend anything like what is going on with Blizzard at EA the computer game company, who's now all the rage in the media because of this anti-woman movement. We don't need to talk about that, that's really over our border. And I don't want to defend that. And I don't want to defend people that are not aware that if you work with women, that you probably can't bring the same jokes you can do just on the men. [...] Of course, some organization like Google or Facebook with all their resources, they should be better. People should be better, but people aren't, I don't have a high opinion of humanity in general, or people in general. I think we should strive to be better. I sound like a Silicon Valley advertising. I hate that, but that's not how I'm made. I think these things will happen, and these things will keep on happening, but hopefully less frequently. This ethical AI group, that Google had, they had it as a PR gag. That's my personal opinion. INTERVIEWER: As a PR gag. What does that mean? PARTICIPANT: Even in the US, privacy gets discussed more and more. And so [Google] can divert some attraction and say, "Hey, we have an ethical group, an ethical AI group, an ethical privacy group, an ethical I have no idea what group, and they use that to look like they are still a nice company out of a garage in Palo Alto. And they don't mean it. They don't have it built into their DNA.}
    
    \item{It's something that's important to me, that we are very concerned about data privacy. We are vehemently against just giving people access to our users' data in a Facebook-esque way, just selling it for profit, et cetera. That's one AI morality line that we've drawn in the sand. [...] As the person responsible for data governance at the company, I wouldn't really accept any other way of doing it. However, I am lucky that all of the other people at the company share my opinion.}
    
    \item{In my opinion, all AI should be done very carefully. Because what you know as the machine learning practitioner, as the data scientists, you know the data that goes into the model, but beyond that, it's not really explainable what the model is doing. It's gotten better over the years, but in a way, most of these giant models are still black boxes. So all you really have control over in terms of trying to eliminate bias or in the modeling process, is the data that you can see as input. [...] Certain features shouldn't be used just in general. And then in terms of privacy, [...] when I started to learn what all these big companies were doing with their data, particularly [data sharing] practices, making it opt out rather than opt in, in terms of what they were allowed to do with your data, that has always frustrated me. Considering all those things and thinking about fairness and how I would want to be treated as a person, I think I should extend the same courtesy to other people. INTERVIEWER: [...] So how do you deal with that tension about your aspirations in terms of things that you might want to do from an ethical perspective in the long-term, versus the things that you need to get done right now? PARTICIPANT: I think it really does come down to the severity of what it would mean to make a mistake in context. The dialects of English example that I gave earlier, what it would mean to make a mistake there, the negative impacts of that are more so on us because the user who our app doesn't work as well as it could for, they'll just be like, "I don't really like this app. Let me go download another app, or let me just not use it." That's totally fine. That's everyone's decision. Obviously I think our app is cool. So they lose out on that. They're not really losing anything in the grander scheme of things, whereas we're incurring the risk on that side because if a user didn't have a good experience on our platform, they didn't think it worked very well, we lost a user. Maybe a more extreme counterexample, there's a cancer diagnosis AI looking at X-rays or something. If you make either a false positives or false negatives in that space, are super negatively impactful to that person. Because if it's a false positive, they get all this chemo that they don't need, and that just makes their life hell. Or if it's a false negative, then something goes unchecked that might potentially doctors could have helped with. In that sense, the user or the person the AI is acting upon, is incurring the risk. And I think that's a distinction.}
    
    \item{We have [talked about ethics]. I would say that it's something we've talked about needing to be a focus of our conversations around growth, but we haven't dug into it deeply at this stage because I think a part of what we'll have to figure out and what we'd like to put on our radar, but haven't gotten too deep on any of it is, what is our role [because we provide a service to other businesses]? I imagine that we'll want a pretty thoughtful framework around that as we grow because an entire industry that we hadn't even thought of, that we may want to have particular considerations for, could come to us and want to start using us. And I think we'll want to be ready with some principles around, oh, well, we hadn't conceived of your use case, but we have some underlying principles that guide how we'll interact with this use case. But no, we haven't actually gotten to those principles just yet.}

    \item{[Ethics is] not something we have talked about. I mean I talk about that a lot at [my main employer], but not in [my startup]. Just because what we're doing isn't approving loans for people, you know what I mean? It's telling people largely insights about their data and not a bunch of other people's data. So it's largely very personal, it's one of the reasons why I have elected to do things in the way we've done. I'm sure there's someone who could probably find some societal impact to what we do, but I don't think it's too great. We're not clustering based on race, ethnicity. We do cluster based on where you live and who else is there, who [is similar to you] and things like that. But I don't really perceive that as something that is messing with society, right?}
    
    \item{[...] the biases, if we don't have the right mechanisms in place as we start this journey and we are not aware of the biases that come into play when we're building out and testing these algorithms, then we're not going to be able to design for all. We'll be designing for white women, for example. And that's not the purpose of what we're trying to accomplish. We're also trying to make sure that, in particular, people of color, women of color who already have biases in healthcare against them, that this avenue, using [Company Name], is a way to help overcome those biases so they can get proper healthcare for themselves.}
    
    \item{But I wanted a diverse team. I want people from different backgrounds because I think that improves not only innovation but it improves those checks and balances from the AI ethics standpoint. [Participant describes specifics of the team members' demographic characteristics and nationalities]. So we all come from different backgrounds and when we're thinking about how do we design this for our target market--women aged 20 to 54 based upon our market research--we want to make sure that we have different perspectives included in how we're designing this for them and for that scale and growth. So I needed people on the team that not only could connect with the customer and empathize with the customer but knew the technology, how to structure it and then that knew how to scale it.} 
    
    \item{Certainly, something that comes up all the time is disrupting jobs, taking jobs away. So in [our] industry, there's actually a labor shortage right now. It has been going on for a long time, particularly around skilled labor. So we frame it in terms of allowing your good people to do higher value things, because you're not going to get rid of your good people, just because a computer can do part of their job. [...] I certainly do talk to my friends and staff about how Facebook is destroying our society and stuff like that. But not so much in the context of these businesses. INTERVIEWER:  It sounds like you've actually considered this quite a bit in terms of the business and its broader implications. Have those conversations translated into anything concrete yet, in the product itself? PARTICIPANT: Well, you have to think about workflow for everyone, throughout the whole process from the guy who's getting paid 15 bucks an hour, all the way through to the executive. And I saw this in medicine too, for sure. Oftentimes, you sign a product for the doctors, but the nurses can kill it, if the nurses don't like using it. So that certainly comes in. In the investor discussions, it comes in quite a lot. You talk to VCs, they'll bring up, "What about this?" [...] And so you've got to deal with the latest headline they saw on the New York Times and be responsive to that. Because they want to know that you're coming from the same place that they're coming from.} 
    
    
    \item{I'm a futurist, so I don't believe in ethics really. Anything that's possible will be made. So all technology is inevitable. It just depends on the timescale you're evaluating. INTERVIEWER: So how does that perspective then inform how you think about building something? PARTICIPANT: We always build to the extent that we can. Whatever is possible, we build for that. We don't limit by current ethical constraints because that's all they are. They're just current ethical constraints. They may not exist in the future.}
    
    \item{I'm on this impact high horse. It's sometimes like I act like I'm still a college student, and naive and not jaded. [I brought up in my company], "Have we thought about going to emerging economies, places where they don't have access to [medical specialists]?" Not just from [a regulatory perspective]. But just that's where the need is the biggest. [But] honestly it's all about, can you stay afloat? [Participant describes company rhetoric about ethical impact.] So that's the ethical stuff that we talk about, but more so for internal propaganda, I think. I'm a little skeptical.}
    
    \item{One thing that we really try to do, is ensure that specifically on [aspects of the underlying science...], that they all have multiple research reports across different populations. So, we're looking at a broader population base, versus the standard genetic reporting is often done on Caucasians in specific areas. [...] And then yes, in terms of the accessibility and democratizing data access, that's one trend that we're really looking at and continuing to focus on.} 
    
    \item{So, we do think about data portability. It is part of our roadmap, but I would not say we're anywhere near a hundred percent on that. We more think about it as interoperability between other products and solutions, that our members will be using, so that it's not siloed with us. Which a lot of companies make it so it is siloed, which makes it more difficult to create the biggest impact for that person. And that's something that we've built our company around, making sure that not only can we use a variety of data sources so that we make it easier for people to get access to our products, but how we can integrate with others so that it's creating an even bigger impact.}
    
    \item{But what I think constitutes AI is a system that is not only capable of performing all the tasks that I described before, doing things consistently and repeatedly, but doing it in an equitable fashion where it's being trained with datasets that are more representative of the world and all of its communities and not just a subgroup. Which is why facial recognition artificial intelligence is so controversial, because the people that are training this AI are being lazy or they're not looking at these datasets of people that don't look like them. If you look at, was it Clearview AI with the police department and stuff, that was just mislabeling, just recognizing, oh, we don't know what a Black person looks like. This person's Black. I think [they committed a crime] and then wrongly convict somebody. So I think artificial intelligence is a wonderful tool, specifically a tool, that needs to be treated and trained with respect and careful diligence.}
    
    \item{We're going to follow the GDPR to some extent, though GDPR is not quite applicable to the U.S. side. We want to use some of the GDPR principles. There's nobody going to stop us building the privacy as tightly as possible into the platform. If somebody has already built a specification around it through the GDPR, or somebody else, the Chinese, they are building some privacy document, we could probably use it. By the time the US puts down the regulations on the privacy aspects, maybe we have built 90\% of it into the platform. So we're going to do everything possible to build the very tight privacy and disclosures.} 
    
    \item{So those are external constraints, and we have to follow those constraints, so data protection, privacy, HIPAA compliance. All of those things we need to build into our product. INTERVIEWER: Do you think about those things as helping your product development or hurting your product development or irrelevant? PARTICIPANT: So the product only exists to serve the customer, and so if the customer is required to follow those regulations, then we must follow those regulations or else we can't serve those customers.} 
    
    \item{[...] And that [product feature] has been trained on data sets that are [details of datasets] and more ethically aligned. So that way we make sure that we're classifying as best as we can. So that way there's no discrimination.}
    
    \item{For me, hits close to home when you hear people are being wrongly convicted of crimes they didn't commit just because an AI said, it was them that did it. I think that my personal relationship to how AI can be mistreated, especially against someone that looks like me, kind of drove me to looking more to the ethics of artificial intelligence and how these models are being trained.}

    \item{There are a lot of people who do not meet that classification [as an accredited investor]. We are very much in a market where people are more educated about investing options. And even though you may not have the assets or capital to do what an accredit investor can, you can still take a portion of your income and make small investments in a lot of companies. [Crowdfunding] gives people more options to invest in companies that are not publicly listed yet. So I think it's more of a play towards what the wider audience that is interested in it and then also in the spirit of transparency and just saying, "Hey, if there's people out there that don't like this, we have a public page that you can express your concerns on that we will address that are on the same page where people make investments." So that way you can get a good picture of where we stand on happening events or things that we've addressed in the past.}
    
    \item{I would say institutional investors want to invest in AI companies that are very mindful of how they're applying AI and doing the thinking for the investors essentially being like, look we are aware of all of the problems at large and upcoming regulation and all of the confusion around it. And we want to stay ahead of it and educate people about it and do it as transparently as possible to make sure that people are comfortable with the solution before it is deployed in mass.}
    
    \item{We're not as big as the giants. Yes it is a David versus Goliath story. Yes I'm not denying it. Of course we're David. We're probably even smaller than David. We're probably toddlers when we consider the competition. But at least we have the gumption and the courage and the guts to think that we could solve a huge problem in the market by using the AI and putting democratization in the hands of people to use [the tools of industry experts].}
    
    \item{Also you can look at my age going into an industry that's very old, very white, so they kind of discount you, write you off, those typical things. In fact, one of the CEOs of a large [company in the industry] is also Black. And we had a call. He was on the call because I was obviously pitching the company to bring them on as a customer. There are so few women and Black people in the [the industry], especially in the forms of leadership and power. [This individual was in] a very unusual position by the standard, which is him being the CEO of this huge company. He reached out to me afterwards and said, "There's [very few] of us. I want to make sure that I see you succeed, because what you're doing is incredible and it's strong but you have to make sure that you play the game. You have to play the game a little bit with people because unfortunately that's just how it is, but I want to make sure I do everything in my power," is this what he said to me, "to make you another successful Black leader in a community that doesn't really reflect us." INTERVIEWER: That's really interesting. What do you think he meant by play the game? PARTICIPANT: Like the good old boys type of people, they're typically pretty conservative or they're [typical workers in entry-level positions], but the higher ups, this is just how they think. And I've seen it. I've spoken to a lot of these people and the way that they think is fundamentally different than the way that me, or I think anyone that's close to me thinks, so you have to kind of dress the part, speak the part, walk a straight line, be careful with who you interact with. It can be a little political, so we just got to make sure you know how to walk the walk. INTERVIEWER: That sounds really difficult. Can you think of examples where you've changed the way you talk or how you're presenting to companies? PARTICIPANT: Oh, definitely all the time. Even my COO who just started, he even notices. It's called code switching.}
    
    \item{Hiring in universities is a big thing. So it's very hard to find women engineers, which is something that we've been looking for and bringing more people that just aren't the norm into positions of high power. So that takes looking at different places for recruiting. You don't look at the Harvards and the Princetons of the world, because you know exactly what you're going to get. You're looking at the Howards and the Spelmans of the world to kind of pull those people. [...] You're even helping in mentoring kids in high school, kind of getting there, kind of catching them a little early, even in college and bringing them up in that light. INTERVIEWER: [...] Howard and Spelman are also elite institutions. Is there a ton of competition then for candidates out of those kinds of pools as well? PARTICIPANT: No. No. INTERVIEWER: Really? PARTICIPANT:Yeah. A lot of it's still posterity from what it feels like. My COO came from [another company in the industry] and you're seeing these career fairs and a lot of these big companies aren't even going there. They're not even looking. If they do, they're not hiring anybody. They're not making the decision to do it. It just seems like just PR.}

    \item{And [a related company], I guess were founded in [a few years ago], but they've done a really good job of democratizing access to data, and helping people understand their data and create some actions around them, that can actually help improve [an aspect of user health]. INTERVIEWER: So, would you say that that's similar to the mission that you're trying to achieve? PARTICIPANT: I don't know that I would say it was similar to the mission necessarily, but democratizing access to data and helping people improve their life, yes. That's similar.}
    
    \item{If you have a model that is more accurate [for white people], then it's not really useful if you have a very diverse crowd of people and considering the fact that we operate worldwide, we've got to make sure that we're covering all nationalities and ethnicities and all that as best as we can.}
    
    \item{[Referring to the security industry] Their whole job is to over extend, to protect the interests of whichever entity that they're trying to protect. And so it's just traditionally a community and school of thought of trying to get as much information as possible instead of just what is necessary to accomplish a job.}
    
    \item{[Regarding privacy] What I said was that people usually have two reactions. Yes, I love it. Take all of my data. I don't care at all. Just make my life easy. Then we have some people that completely freak out and they get very, very scared. We like being in the middle. We don't want to be [at one extreme]. We don't want to be [at the other extreme] either. There are more people who are worried than people who are not worried based on statistics, on what we have seen. I'm always trying to understand, what are you really afraid of? I think the fear comes from a couple of things. One, they don't understand how the technology works. When you don't know how something works, you are afraid of it. It's default. When humans don't understand something, we are scared of it. This is a very useful biological trait in trying to protect ourselves from the unknown. I think that's what it boils down to. The second thing is that they feel a lack of control, which is very fair. Very fair. What I'm always saying is you shouldn't be afraid. You should be concerned in the following sense. If you educate yourself, you know the exact risks, the dangers that come with any technology, and then you find ways of addressing and mitigating it, but you have to be informed and it's hard to inform the public. If you're trying to educate the market, they have estimated that it takes about \$200 million to educate the market. Lots of marketing and education meant for promotion. We couldn't do that alone. We didn't see a lot of other companies implementing things with the right safeguards. We said, okay, what is the next thing we can go to so that we don't have to worry about this? INTERVIEWER: What do you think are the right safeguards? [...] PARTICIPANT: Multiple ones. First of all, the default has to be opt out. You can have the default opt in and then people go back to remove data. That's the first thing. Second thing, retention. Don't keep [data] more than you need to. You know what you need to have another biometric enrollment [...]. It's not very hard to do. Why keep it forever? That thing is giving control to people. If you want to delete an account, you don't know if they information is actually deleted on the servers versus just deactivated. People need to have control over that. It has been very clear and it's not hard to do. It's really not hard to do. Implement it. It is the lack of will in most cases or concern about people's information. Once you put all of these things together, you have already reached a very, very good threshold of security. It will never be perfect. It will never be perfect, but it'll be much better than what we have right now.} 
    
    \item{But for us, we are getting more determined to be the ones that will lead because we genuinely care. Plain and simple. Everybody in our team, we genuinely care. I know we will not get everything right. I know we will make mistakes. I know we will have backlash. I can see all of these things, but when you want to be leading in anything and you want to drive progress, you have to be willing to take the punches. That's our approach. But going back to your original question, the only concern we have on this one, the biggest concern I would say is that [companies who over promise and under deliver] are diluting the market in the following sense. If you're a decision maker, let's say you are the CMO of Coca-Cola. You really care about privacy and you want to select the best company and you really care about the results and you make the investment. If you work with a vendor that they're not so good and they offer promise and you get PR nightmare or whatever, it's very unlikely you're going to try again. They are delaying innovation. They are delaying innovation that makes it harder to get good solutions out to the market.}
    
    \item{I think to some degree [beliefs about privacy] can be generational. There's a shift towards privacy in younger demographics, in technology. There's another thing that has happened that most of the people, I don't know if I'll say our age or younger, they're starting their own businesses or they're getting to the C level positions. They were born and raised with phones in their hands, with technology in their hands. They are more receptive.}
    
    \item{We have investments from private individuals, and while we were trying to raise funding, there were some investors that they really cared about the fact that we are trying to take an ethical approach in what we are doing. They really, really like that and supported that. There were some other investors that were turned off by it. You see both. I heard by at least one person saying it out loud and more than one implying it, saying, "give [your product away] for free and just get all the data and sell all the data," which is not what we want to do, but this exists. For the most part, we are trying to raise capital from investors that have the same kind of values and mindset with us and people who are not afraid to lose certain revenue or sales just to follow the same values. We had clients asking us to do things that we said, you know what? No. No, this is not something we feel comfortable with doing.}
    
    \item{[When evaluating service providers] I think a lot about equity and promotion of different people, people that don't fit the "norm." So like women founders, like Black, Latin, and brown people that are founders of startups, making sure that they have a strong mission, and also have a great system or platform or product. And so the [data labeling] company that we're working with came as a recommendation from a friend of mine who actually looked up their work. They have great customer success rates, great customer success stories. Very, very large companies utilize them, and they're equitable not only because [the company] was founded and started by a woman, but they also went a step further. It really solidified [the partnership] in my mind that they went to different parts of Africa where they teach individuals that wouldn't typically have access to these types of these jobs, how to do these types of jobs, and [this company] actually employs them.}
\end{itemize}

\bibliographystyle{ACM-Reference-Format}


\end{document}